\documentclass[pra,twocolumn,preprintnumbers,amsmath,amssymb,superscriptaddress,showpacs,longbibliography]{revtex4-1}
\usepackage{graphicx}
\usepackage{amsmath}
\usepackage{graphics}
\usepackage{amssymb}
\usepackage{amsfonts,epsfig}
\usepackage{natbib}
\usepackage{hyperref}
\usepackage{enumitem}
\usepackage{color}
\usepackage{ulem} 

\newcommand{\ket}[1]{\left|#1\right>}

\newcommand{\beq}{\begin{equation}}
\newcommand{\eeq}{\end{equation}}
\newcommand{\bea}{\begin{eqnarray}}
\newcommand{\eea}{\end{eqnarray}}

\newcommand{\mom}{\mathbf{m}}
\newcommand{\mean}[1]{\langle{#1}\rangle{}}

\newcommand{\vG}[1]{\mathbf{g}^{(#1)}} 				
\newcommand{\vnz}[1]{\mathbf{n}_z^{(#1)}}			
\newcommand{\vR}[1]{\mathbf{r}^{(#1)}}				
\newcommand{\rot}{\mathbf{R}}							
\newcommand{\proj}{\mathbf{P}}							
\newcommand{\vB}{\mathbf{b}_0}							
\newcommand{\attfc}[1]{\eta_{#1}}							
\newcommand{\rechi}{\eta_{\{ 12 \}}}
\newcommand{\imchi}{ \eta_{[12]} }
\newcommand{\dphi}[1]{ \omega_s \Delta \tau_{#1} } 	
\newcommand{\shift}{\phi }									
\newcommand{\cz}[1]
							{#1}
\newcommand{\nb}[1]
							{#1}
\newcommand{\zi}[1]
							{#1}
\newcommand{\out}[1]
							{}

\begin{document}

\title{Localization of a magnetic moment using a two-qubit probe}
\author{Jan Krzywda}
\affiliation{Institute of Theoretical Physics,
Faculty of Physics, University of Warsaw, ul. Pasteura 5, PL--02--093 Warsaw, Poland}
\affiliation{Institute of Physics, Polish Academy of Sciences, al.~Lotnik{\'o}w 32/46, PL 02-668 Warsaw, Poland}
\author{{\L}ukasz Cywi{\'n}ski}
\email{lcyw@ifpan.edu.pl}
\affiliation{Institute of Physics, Polish Academy of Sciences, al.~Lotnik{\'o}w 32/46, PL 02-668 Warsaw, Poland}
\author{Piotr Sza\'{n}kowski}
\email{piotr.szankowski@ifpan.edu.pl}
\affiliation{Institute of Physics, Polish Academy of Sciences, al.~Lotnik{\'o}w 32/46, PL 02-668 Warsaw, Poland}

\date{\today}

\begin{abstract}
A nanomagnet precessing in an external magnetic field can be treated as a source of narrow-bandwidth magnetic noise, that leaves characteristic fingerprints in decoherence of a nearby spin qubit undergoing dynamical decoupling. We show how, by measurements of two-qubit coherence, a noise sensor composed of qubit pair can be used to reconstruct the position of the nanomagnet. Such localization of noise source is possible with only two qubit probes, because the course of coherence decay under appropriately designed dynamical decoupling sequences contain information not only about noises experienced by each qubit, but also about their cross-correlations. We test the applicability of the proposed protocol on an example of two qubits coupled to the nanomagnet via dipolar interaction. We also show how, using a two-qubit sensor possessing a particular symmetry, one can localize the nanomagnet even when the sensor-magnet coupling law is unknown.
\end{abstract}


\maketitle

\section{Introduction}
Qubits can be used as sensors of environmental noise and/or signals \cite{Degen_RMP17,Szankowski_JPCM17}. Probes made out of single qubit have been used to characterize environmental noise spectra in a variety of settings \cite{Biercuk_Nature09,Bylander_NP11,Alvarez_PRL11,Yuge_PRL11,Kotler_Nature11,Almog_JPB11,Medford_PRL12,Dial_PRL13,Muhonen_NN14,Romach_PRL15,Malinowski_PRL17}. Such a spectral reconstruction is made possible by applying a dynamical decoupling (DD) sequences of unitary operations that correspond to periodic ``flipping'' of the qubit state with virtually instantaneous $\pi$-pulses \cite{Viola_PRL99,Uhrig_PRL07,Gordon_PRL08,Cywinski_PRB08,Paz_NJP16}. Such sequences act effectively as narrow-band pass filters with characteristic frequencies determined by the number of pulses and the magnitude of intervals between them, see \cite{Szankowski_JPCM17} and references therein. In the case of an environment that consists of localized sources emitting noise with narrow spectra, e.g. nuclear or electron spins precessing in external magnetic field with frequency determined by its value and their gyromagnetic factors, a qubit becomes a probe allowing for magnetic resonance imaging (MRI) with nanoscale resolution \cite{Staudacher_Science13,DeVience_NN15,Haberle_NN15,Wrachtrup_JMR16,Degen_RMP17,Lovchinsky_Science16}. 

In nanoscale MRI context one can try to obtain information about the number of spins with given precession frequency that are in the vicinity of the qubit \cite{Staudacher_Science13,Degen_RMP17}. Alternatively, one can try to use the qubit to precisely localize the source of signal of specific frequency. While sensing of single nuclear spins \cite{Zhao_NN12,Ma_PRAPL16}  and small nuclear spin clusters \cite{Zhao_NN11} often requires a quantum mechanical treatment of their interaction with the qubit, sensing of a group of nuclei that are localized in space, e.g.~a molecule positioned at some distance from the qubit \cite{Lovchinsky_Science16}, can be modeled by treating the spins as a classical nanomagnet precessing in external magnetic field, see \cite{Szankowski_JPCM17} for an example calculation. In the presence of many such nanomagnets in the vicinity of the qubit, focusing on the signal coming from one of them is made possible by the application of a magnetic field gradient \cite{Arai_NN15,Zhang_arXiv16}, which is a basic tool of MRI. Adjusting the period of the DD sequence applied to the qubit one can make it sensitive to environmental noise/signal at a set of discrete frequencies that correspond to a set of surfaces of constant magnetic field, making the qubit sensitive to a subset of the nearby sources. In this paper we focus on the source that is treated as a classical magnetic moment, and we propose a way of finding its location by using coherence measurements on two nearby qubits.

For a given form of qubit-source coupling law, such as dipolar coupling in the case of nitrogen-vacancy (NV) centers \cite{Doherty_PR13,Dobrovitski_ARCMP13} used for localization of nuclear or electron spins \cite{Maletinsky_NN12,Staudacher_Science13,Rondin_RPP14,DeVience_NN15,Haberle_NN15,vanderSar_NC15,Wrachtrup_JMR16,Degen_RMP17}, measurement of decoherence of a {\it single} qubit allows for finding of a two-dimensional surface of possible source locations. More detectors are needed to further pin-point the location of the source: in three dimensions three detectors are needed to perform such a {\it trilateration} of the source position \cite{Ma_trilateration_PRAPL16}. 
Separate decoherence measurements of two qubits that are set apart in space, single out a curve (a one-dimensional surface) on which the source is located. While adding a third qubit to achieve classical trilateration is an obvious solution, it might not be the most practical one after taking into account that well-working qubits are still rather precious resources. 

Here, we show how adding measurement of one of two-qubit coherences can be used to find an additional surface to complete a traditional trilateration protocol with two qubits only. This follows from the observation that decay of certain two-qubit coherences is sensitive to the presence of cross-correlations of signals affecting each qubit. In fact, as discussed recently  in  \cite{Szankowski_PRA16}, using distinct DD sequences applied to each qubit, one can fully characterize the real and imaginary part of the spectrum of cross-correlation.
In the protocol discussed in this paper,  the necessary inputs are two measurements of single-qubit decoherence (each qubit subjected to a periodic DD sequence having its period adjusted to characteristic frequency of the source), and a measurement of one of two-qubit coherences that can be reconstructed from a subset of two-qubit tomographic measurements. The practical implementation of such protocol is significantly aided by the possibility of using two sequences of pulses -- each applied to another qubit -- that are shifted in time with respect to each other (note that this requires the two qubits to be addressed separately, which is easy for NV centers located in a magnetic field gradient, or having non-parallel quantization axes). 
As an example, we simulate the performance of the protocol in the case of two qubits coupled via dipolar interaction to the source. We also show how one can exploit the symmetry of the two-qubit sensor and the ability to change the relative position of the sensor and the source to localize the latter even when coupling between the two has an unknown form.

The paper is organized in the following way. In Section \ref{sec:model} we describe the model that we focus on: two qubits with possibly disctinct quantization axes, each coupled longitudinally to a precessing magnetic moment, that is described as a source of classical Gaussian noise. In Section \ref{sec:decoherence} we present the solution to the problem of decoherence of two qubits that are coupled to the previously described noise source, and subjected to two (possibly distinct) dynamical decoupling sequences. This solution is then used in Section \ref{sec:protocol}, where we present two versions of  the source localization protocol: in the first we assume that the relative position of the two qubits is known, while in the second we assume that it is unknown. Finally, in Section \ref{sec:symmetry} we show how a sensor possessing a particular symmetry can be used to localize the source, even if the source-qubit coupling law is unknown. The Appendix contains details concerning the simulation of the application of the protocol to the case of qubits interacting with the source via dipolar interaction.

\section{The model} \label{sec:model}
\subsection{The noise source} \label{sec:source}
We model the source of the oscillatory  magnetic field affecting the qubits as a classical magnetic dipole $\mom(t)$. We assume that $|\mathbf{m}(t)|$ is constant on the time scale of a single experiment run, but changes in an uncontrollable manner from run to run, i.e.~the initial value of $\mom(0)$ is treated as a random variable. This corresponds to the situation in which the dipolar interactions between the spins forming the magnetic moment and spin-lattice interactions, responsible for relaxation of these spins, operate on timescales much longer than the single experimental run.
Then, the source evolves only by precessing with known angular frequency $\omega_{s}$ due to the presence of a given external magnetic field $\vB$. Note that the above assumptions mean that the magnetic moment is a quasi-static random variable in the reference frame rotating with frequency $\omega_s$.
The time dependence of $\mom(t)$ in the laboratory frame, where the $z$-axis is parallel to $\vB$, is given by
\begin{align}
\mom(t)&=\left( \begin{array}{c}
m_x(0)\cos\omega_st -m_y(0)\sin\omega_s t\\[.1cm]
m_x(0)\sin\omega_st + m_y(0)\cos\omega_s t\\[.1cm]
m_z(0) \\
\end{array}\right)\nonumber\\[.25cm]
&=\rot(\omega_s t)\mom(0)\,.
\end{align}
Here, $\rot(\omega_s t)$ is the matrix of rotation by angle $\omega_s t$ about axis parallel to $\vB$. We assume that the distribution of $\mom(0)$ is Gaussian and isotropic, therefore given by
\beq
P(\mom(0)) = \frac{1}{(\sqrt{2\pi}\sigma)^3}e^{-\frac{1}{2\sigma^2}\mom(0)\cdot\mom(0)} \,\, .\label{eq:mom_dist}
\eeq
The necessary condition for approximate equivalence between the above classical treatment of the magnetic moment and the exact quantum treatment is for the moment to be composed of $N \! \gg \! 1$ spins of length $J$. The distribution is isotropic when the density matrix of the constituent spins is maximally mixed, as it is the case for $k_{B}T \gg \hbar\omega_{s}$---the conditions typical for nuclear spin systems at experimentally realistic temperatures. This makes the classical model potentially applicable to localization of proteins containing many nuclear moments by NV centers \cite{Staudacher_Science13,Lovchinsky_Science16}. In technical terms, when this classical approximation holds, in calculation of the evolution of the qubit-probe, one can replace tracing over the density matrix of the spins with averaging over $\mom$ distributed according to (\ref{eq:mom_dist}). Then, for the description to be consistent, the rms must be related to the microscopic parameters associated with the signal source in a following way \cite{Merkulov_PRB02}
\begin{equation}
\sigma = \sqrt{\frac{1}{3}J(J+1)N \mu_{J}} \,\, , \label{eq:sigma}
\end{equation}  
where $\mu_J$ is the magnetic moment of spin $J$.

An additional requirement for the classical approximation to be valid is to have a negligible back-action from the qubit-probe, so that the precession of the moment is not disturbed by its presence. In \cite{Szankowski_JPCM17} (see also \cite{Ma_PRAPL16}) it was shown that in the context of qubits subjected to dynamical decoupling, 
the back-action is negligible when the number of applied pulses satisfies $n < \omega/A\sqrt{N}$, where $A$ is the typical coupling of transverse components of single spin $J$ to the qubit. Clearly $\omega/A \! \gg \! 1$ is necessary, i.e.~individual spins comprising $\mathbf{m}$ have to be weakly coupled to the probing qubits. 

\subsection{Sensor-source coupling}
Each of the qubits (labeled by $q=1,2$) is assumed to have a well-defined quantization axis $\mathbf{n}_{z}^{(q)}$, so that their self-Hamiltonians are of the form $\hat H_q=\Omega_{q}\hat{S}_{z}^{(q)}$, with $\hat S_z^{(q)}=(1/2)\hat\sigma_z^{(q)}$ being the spin-$1/2$ operator component parallel to $\mathbf{n}_{z}^{(q)}$. For spin-$1/2$ qubits (e.g. spins of electrons localized in quantum dots or on donors), the axes are determined by local magnetic fields. However, in the case of qubits based on NV centers, which are currently the most widely investigated in the context of noise/signal sensing, the axes (at least for typically used values of $\vB$ fields) are determined by the geometry of the nitrogen-vacancy complex, specifically by the direction of vector connecting the nitrogen impurity with the vacancy \cite{Doherty_PR13,Rondin_RPP14}. This is due to the fact that the qubit is based on $m_{s}=0$ and $1$ levels of spin-$1$ electronic complex, and the crystal field splitting, that is large enough to dominate over typically used external fields, sets the quantization axis to be parallel to the vector connecting the nitrogen impurity with the carbon vacancy.

An important assumption is that the magnitudes of the fields experienced by the qubits due to the presence of the magnetic moment $\mom$ in their vicinity are much smaller than their splittings $\Omega_{q}$. Consequently, the components of these fields that are perpendicular to $\mathbf{n}_{z}^{(q)}$ can be ignored, and only the longitudinal coupling of the qubits with the components of $\mom(t)$ is taken into account. This leads to the {\it pure dephasing} Hamiltonian of qubit-source interaction: 
\beq
\hat H_{\text{int}} = \sum_{q=1,2} \hat S_{z}^{(q)} \left( \vG{q} \cdot \mom(t) \right)\,\, ,
\label{eq:Hint} 
\eeq
where the vector $\vG{q}(\mathbf{r}^{(q)})$ represents the qubit-source coupling that depends on the vector $\mathbf{r}^{(q)}$ connecting the qubit $q$ with the point-like source. 

\begin{figure}[tb]
	\includegraphics[width=0.5\textwidth=]{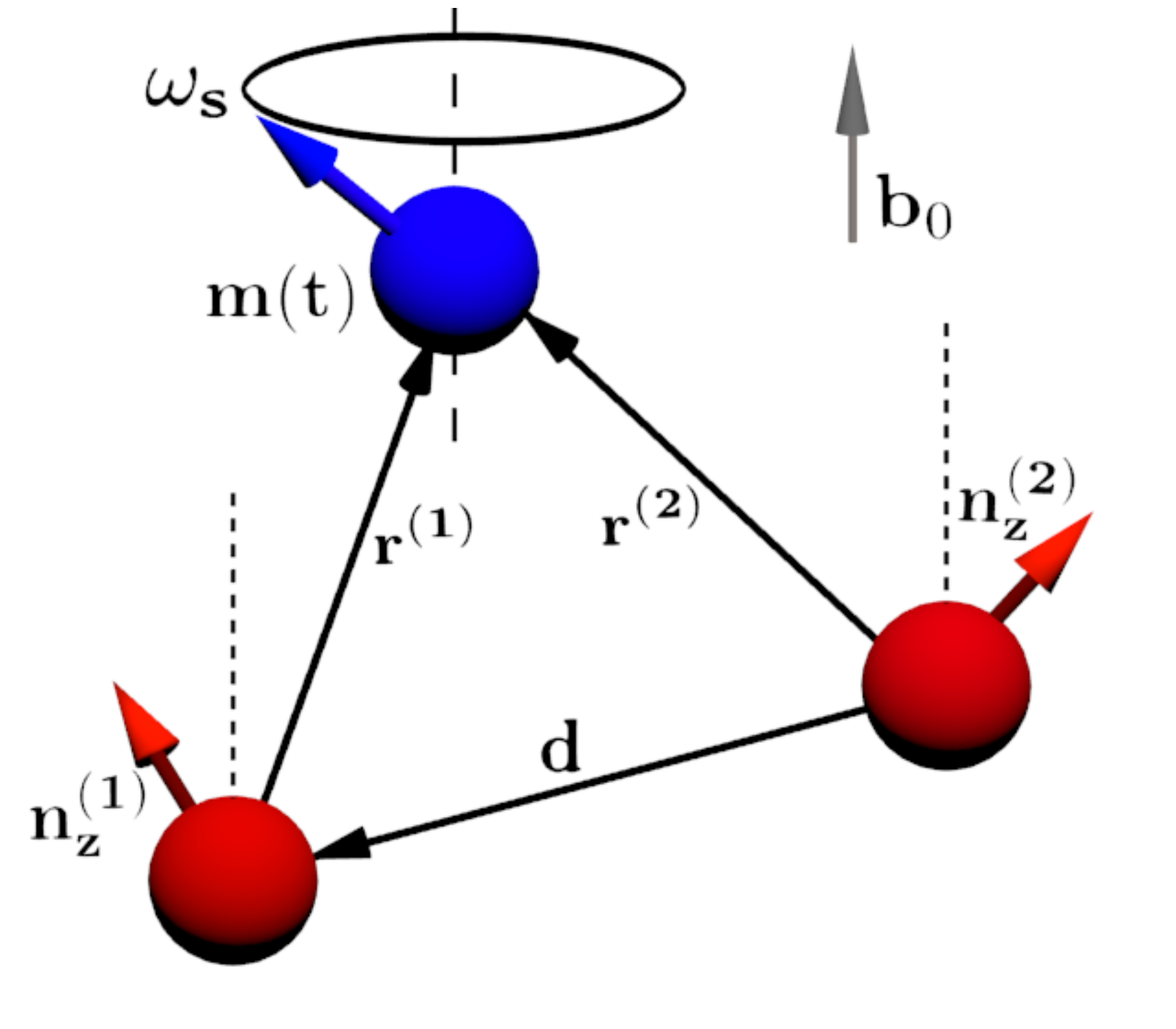}
	\caption{The schematic diagram of the system: $\mathbf{d}$ -- the qubit-qubit displacement vector, $\vnz{q}$ -- the unit vector pointing in the direction of the quantization axis of qubit $q$, $\vR{q}$ -- the qubit-source displacement vector, and $\vB$ -- the applied control magnetic field. The source, modeled by localized classical magnetic moment $\mom(t)$, precesses around field $\vB$ with known angular frequency $\omega_s$. The initial value of $\mom$ changes randomly between sensor coherence measurements according to isotropic distribution~\eqref{eq:mom_dist}.}
	\label{fig:system}
\end{figure}

A schematic diagram of the sensor-source configuration  is presented in Fig.~\ref{fig:system}. Note that in the most general case shown there, neither of the qubits' quantization axes $\mathbf{n}_{z}^{(q)}$ is parallel to the external field $\vB$. The vector $\mathbf{d}$ connecting the two qubits comprising the sensor should be known for the ease of implementation of the localization protocol (with tools such as atomic force microscope that was used to determine the locations of nearby NV centers in \cite{Grinolds_NP11}), but as we show below, localization of the source can be achieved even if $\mathbf{d}$ is unknown at the beginning of the experiment.

The most natural example of qubit-source coupling is the dipolar interaction, for which
\begin{equation}
\label{eq:dipole_int}
\vG{q}_\mathrm{dipole} \propto 
	\frac{
		|\vR{q}|^2 \vnz{q}
		-3(\vnz{q}\cdot \vR{q}) \vR{q}
	}{|\vR{q}|^5}\,.
\end{equation}
The localization method presented below can be applied to other types of couplings (e.g.~various kinds of exchange interactions, possibly mediated by free carriers in order to have long range), although the amount of information on the source position that can be extracted and the ease of its extraction will depend on the specific form of the interaction, especially on the degree of its symmetry (see the remarks later on the case of isotropic interaction).

\section{Decoherence under noise filtering}  \label{sec:decoherence}

\subsection{Pulse sequences as a signal frequency filters}  \label{sec:pulses}

Within a single run of the experiment (comprising intialization, evolution, and readout) the evolution of the two-qubit state under the influence of source-qubit interaction \eqref{eq:Hint} is governed by a unitary operator $\hat U(T,0) = \hat U_\mathrm{int}^{(1)}(T,0)\hat U_\mathrm{int}^{(2)}(T,0)$, with 
\begin{equation}
\hat U_\mathrm{int}^{(q)}(T,0) = e^{-i \hat S_z^{(q)} \vG{q}\cdot\big( \int_0^T \mom(t)\mathrm{d}t \big)}\,,
\end{equation}
acting on qubit $q$ subspace only. The application of $\pi$-pulses at times $t^{(q)}_k$ interrupts the phase evolution with the rotation of qubit's Bloch vector, $\exp[-i \pi\hat S_x^{(q)}] = -i\hat\sigma_x^{(q)}$ for $\pi$ rotations about the $x$ axis (we assume ideal pulses, so that the choice of rotation axes is irrelevant). It is important to note that the pulse application times $t^{(q)}_k$ can be {\it different} for each qubit. In fact, as it will be described later, the ability to apply different pulse sequences to each of the qubits is a very useful tool for localization protocols. 

Assuming that there are $n_q$ pulses in the sequence applied to qubit $q$, and $t^{(q)}_0=0$, $t^{(q)}_{n_q+1}=T$, the evolution operators are now modified (see e.g.~\cite{Szankowski_JPCM17} for step-by-step derivation)
\begin{align}
\nonumber
&\hat U^{(q)}(T) = \\
&(-i\hat\sigma_x^{(q)})^{n_q}\exp\left(-i \hat S_z^{(q)}\vG{q}\!\cdot\! \int_{0}^{T}\!\!\! f_T^{(q)}(t)\mom(t)\mathrm{d}t\right).\label{eq:evo_op}
\end{align}
where we encounter the so-called {\it time-domain filter function} \cite{deSousa_TAP09,Cywinski_PRB08,Biercuk_JPB11,Szankowski_JPCM17}
\begin{equation}
f_T^{(q)}(t) = \sum_{k=0}^{n_q} (-1)^k\, \Theta\left(t-t_k^{(q)}\right)\Theta\left(t_{k+1}^{(q)} -t\right)\,,
\end{equation}
and the Heaviside step function is defined as
\begin{equation}
\Theta(t) = \left\{\begin{array}{lcl} 1 &\text{for}&t>0\\0 &\text{for}& t<0\\\end{array}\right.\,.
\end{equation}
Note that in case of evolution controlled by sequences of {$n_q\in\mathrm{even}$} pulses, the eigenstates of {$\hat S_z$} return to the original state after {$n_q$} flips, {$|\pm \tfrac{1}{2}\rangle\to |\pm\tfrac{1}{2}\rangle$}, while for {$n_q\in\mathrm{odd}$} the states are flipped at the end, {$|\pm \tfrac{1}{2}\rangle\to  |\mp \tfrac{1}{2}\rangle$}. It is customary to append the odd {$n_q$} sequences with one more pulse at the end of the evolution in order to avoid the final reversal of states.

In order to demonstrate the frequency filtering aspect of the pulse sequences we analyze the time dependent part of Eq.~\eqref{eq:evo_op}. In the reference frame where the $z$-axis aligns with $\vB$, the integral of filter function and the source magnetic moment reads
\begin{align}
&\int_0^T f^{(q)}_T(t) \mom(t) \mathrm{d}t=\nonumber\\
&\int_0^T \mathrm{d}t  f^{(q)}_T(t) \left( \begin{array}{c}
m_x(0)\cos\omega_st -m_y(0)\sin\omega_s t\\
m_x(0)\sin\omega_st + m_y(0)\cos\omega_s t\\
m_z(0) \\
\end{array}\right)=\nonumber\\
& |\tilde{f}^{(q)}_T(\omega_s)|\left(\begin{array}{c}
m_x(0)\cos\phi^{(q)}(\omega_s)  -m_y(0)\sin\phi^{(q)}(\omega_s)\\
m_x(0)\sin\phi^{(q)}(\omega_s) + m_y(0) \cos\phi^{(q)}(\omega_s)\\
0 \\
\end{array}\right)\,,\label{eq:filter}
\end{align}
where $\tilde{f}_T^{(q)}(\omega_s) = | \tilde{f}^{(q)}_T(\omega_s) |e^{-i\phi^{(q)}(\omega_s)}$ is the Fourier transform of the filter function calculated at $\omega=\omega_s$. In this context it is convenient to express $\tilde{f}_T^{(q)}(\omega)$ in terms of the Fourier series representation of the filter function, 
\begin{align}
&f_T^{(q)}(t) = \Theta\left(T-t\right)\Theta\left(t\right)\sum_{k}c_{k\omega_q}^{(q)} e^{i k \omega_q t}\,,\\
&c_\omega  = \frac{1}{T}\int_0^T e^{-i \omega t}f_T(t) \mathrm{d} t\,.
\end{align}
Here $\omega_q$ is the smallest multiple of $2\pi/T$ present in the expansion; it defines the characteristic frequency of the sequence. Then,
\begin{align}
&\tilde{f}_T^{(q)}(\omega_s) =\int_{-\infty}^\infty f_T^{(q)}(t)e^{-i\omega_s t}\mathrm{d}t =\int_0^T f_T^{(q)}(t)e^{-i\omega_s t}\mathrm{d}t\nonumber\\
&=\int_0^T \left(\sum_{k}c^{(q)}_{k\omega_q} e^{i k \omega_q t}\right)e^{-i\omega_s t}\mathrm{d}t \nonumber\\
&=T\sum_{k}c_{k\omega_q}^{(q)}e^{i\frac{T}{2}(\omega_s-k \omega_q)} \mathrm{sinc}\frac{T(\omega_s-k \omega_q)}{2}\,. \
\label{eq:detune}
\end{align}

\begin{figure}[tb]
	\includegraphics[width=0.5\textwidth]{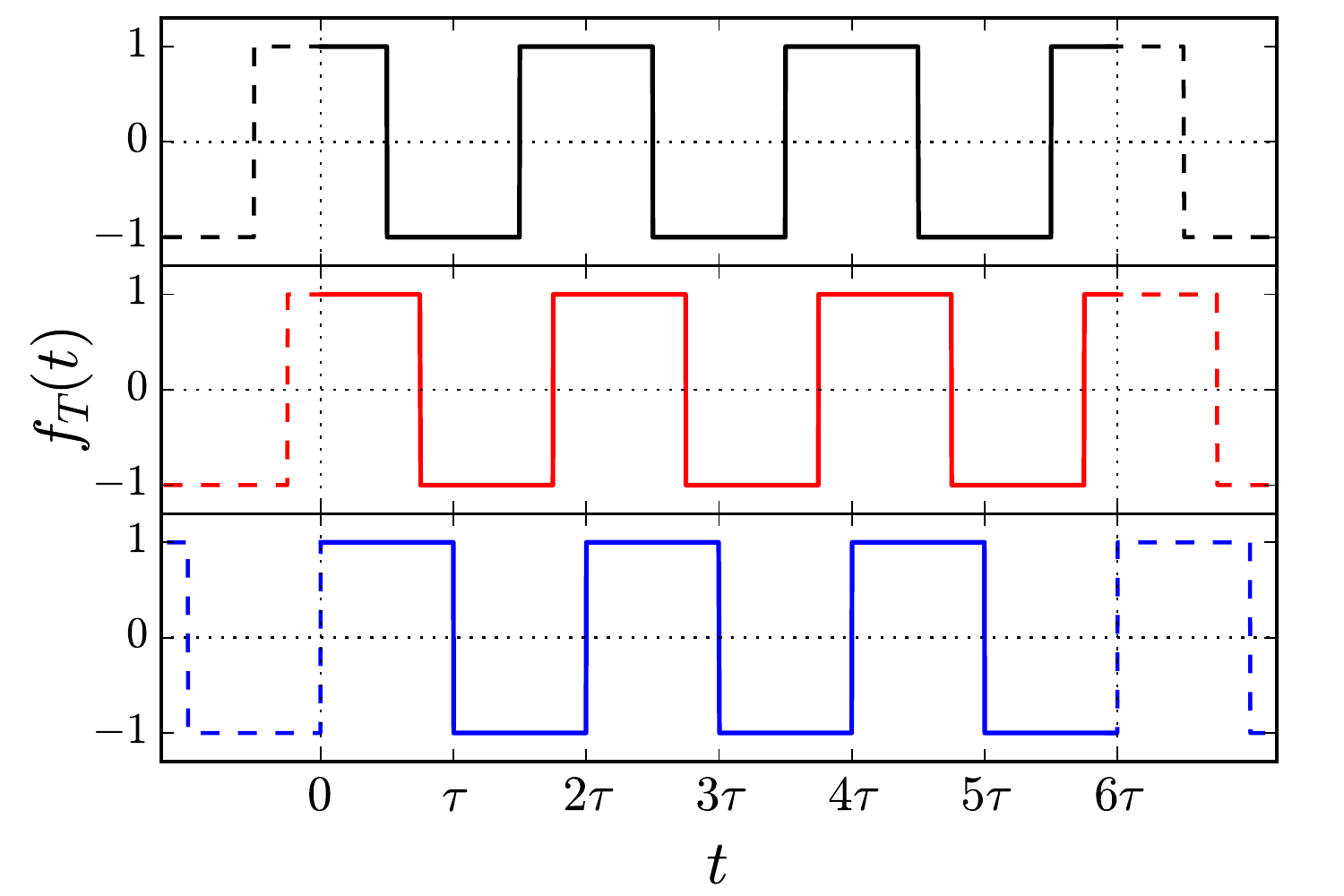}
	\caption{Examples of shifted CP sequences with characteristic frequency $\omega_p =\pi/\tau$ and total duration $T=6\tau$. The solid lines show filtering functions $f_T(t)$ of sequences, while dashed lines indicate periodically extended versions of each filter, $f_{T\mathrm{-period}}(t)$. The top panel presents the original CP sequence, the panel below shows sequence shifted in respect to CP by $\Delta\tau=\tau/4$, and the bottommost panel is a PDD sequence for which $\Delta\tau = \tau/2$.}\label{fig:SCP}
\end{figure}

For the purpose of source localization, the most suitable choice are the sequences belonging to the class of Shifted Carr-Purcell (SCP) sequences. This class is built upon the original Carr-Purcell (CP) sequence defined by pulse times $\frac{\tau}{2},\frac{3\tau}{2},\frac{5\tau}{2},\ldots,\frac{n\tau}{2}$ (and $n\in \mathrm{even}$). In this case the characteristic frequency is determined by the inter-pulse interval, $\omega_q = \pi/\tau$, and the Fourier coefficients are $c^{(\mathrm{CP})}_{k\pi/\tau} = 2e^{i k\frac{\pi}{2}}/(i k \pi)$ for $k\in \mathrm{odd}$ and zero otherwise. The other members of the class are the sequences {\it time-shifted} with respect to CP. We say that one sequence is time-shifted with respect to the other when the corresponding filter functions $f_T(t)$ and $f_T'(t)$ are such that their periodically extended versions $f_{T\mathrm{-period}}(t)=\sum_m \Theta((m+1)T-t)\Theta(t-m T)f_T(t-m T)$ (and analogically for $f_T'$) can be transformed one into another by translation, i.e. there exists $\Delta\tau$ such that $f_{T\mathrm{-period}}(t) = f_{T\mathrm{-period}}'(t-\Delta\tau)$. Figure~\ref{fig:SCP} illustrates a number of examples of such sequences. The Fourier coefficients of time-shifted sequences differ only by a phase factor, i.e. $c_{\omega} = c_{\omega}' e^{i \omega \Delta \tau}$. For example, Periodic Dynamical Decoupling (PDD) sequence, defined by pulse times $\tau,2\tau,3\tau,\ldots,(n-1)\tau$, is time-shifted with respect to CP by $\Delta\tau=\tau/2$, so that $c^{(\mathrm{PDD})}_{k\pi/\tau} = c^{(\mathrm{CP})}_{k\pi/\tau}e^{-i k \frac{\pi}{2}}$.

From this point we will assume that both sequences applied to qubits belong to SCP class and, unless stated otherwise, that their characteristic frequencies are tuned to the precession frequency, i.e.~$\omega_q = \omega_s$, so that the inter-pulse interval is $\tau\! =\! \pi/\omega_s$. The evolution time $T$ is then a multiple of $\tau$, and the source-induced decoherence is maximized. Using Eq.~(\ref{eq:detune}) we obtain
\begin{equation}
|\tilde{f}_T^{(q)}(\omega_s)| = \frac{2T}{\pi}\ ,\ \phi^{(q)}(\omega_s)= \dphi{q}\,,
\end{equation}
with the time-shift $\Delta\tau_q$ defined with respect to CP sequence. If the sequences were not in tune, then Eq.~\eqref{eq:detune} would define the error introduced be the detuning between source frequency and the frequency of the sequence. Note that $(T/2) \mathrm{sinc} [T(\omega_s-k\omega_p)/2]$ tends to $\pi \delta(\omega_s-k\omega_p)$ as $T\to\infty$ (equivalently, as $n\to\infty$), so that in the limit of large $T$ the whole expression vanishes unless the detuning is exactly zero. On the other hand, if the qubits are coupled to other magnetic moments precessing with frequencies different than $\omega_s$, the pulse sequences would decouple the sensor from all signals except for the targeted source. An example of expected dependence of observed coherence on detuning of $\omega_q$ from $\omega_s$ is shown in Fig.~\ref{fig:dip}.

Going back to Eq.~\eqref{eq:filter}, the action of the tuned pulse sequence on precessing magnetic moment can be decomposed as follows
\begin{align}
&\frac{2T}{\pi}\left(\begin{array}{c}
m_x(0)\cos\dphi{q}  -m_y(0)\sin\dphi{q}\\
m_x(0)\sin\dphi{q} + m_y(0) \cos\dphi{q}\\
0 \\
\end{array}\right)\,
=\nonumber\\
& 
\frac{2T}{\pi}
\underbrace{\left(\begin{array}{ccc}
1 & 0 & 0\\
0 & 1 & 0\\
0 & 0 & 0\\
\end{array}\right)}_{\equiv\proj}
\underbrace{\left(\begin{array}{ccc}
\cos\dphi{q} & -\sin\dphi{q} & 0 \\
\sin\dphi{q} & \cos\dphi{q} & 0 \\
0 & 0 & 1 \\
\end{array}\right)}_{\equiv \rot(\dphi{q})}
\left(\begin{array}{c}
m_x(0) \\ m_y(0) \\ m_z(0) \\
\end{array}\right)\label{eq:R}
\end{align}
where $\rot(\dphi{q})$ is the rotation around the applied field $\vB$ by angle $\dphi{q}$, and $\proj$ is the projection onto plane perpendicular to $\vB$.

\subsection{Local and non-local attenuation factors}

The information about source location is to be extracted from the two-qubit state evolving under action of $\hat U(T,0)= \hat U^{(1)}(T,0)\hat U^{(2)}(T,0)$ and averaged over the distribution of the initial values of $\mom$. Under the {\it resonance condition} of characteristic frequency $\omega_q$ of both DD sequences matching the source frequency $\omega_s$, the evolution of elements of the sensor density matrix -- the {\it coherences} -- is then given by (for results in the non-resonant case see Sec.~\ref{sec:spectroscopy}),
\begin{align}
&\rho_{s_1s_2, s_1's_2'}(T) = \langle s_1s_2 |\langle \hat U(T,0) \hat \rho(0) \hat U^\dagger(T,0) \rangle| s_1's_2'\rangle \nonumber\\
&= \rho_{s_1s_2,s_1's_2'}(0) \times\nonumber\\
&\left\langle\exp\left[ -i \sum_{q}(s_q-s_q') \left(\rot(-\dphi{q})\proj\vG{q}\right)\cdot\mom(0)\right]\right\rangle \label{eq:evolution}
\end{align}
where $|s_1 s_2\rangle$ ($s_q=\pm 1/2$) are the elements of the product basis of the eigenstates of qubit spin operators. The average, $\langle\ldots\rangle$, can be carried out analytically using the property of multi-variable Gaussian distributions,  $\langle e^{-i \mathbf{a}\cdot\mom(0)}\rangle = e^{-\frac{1}{2}\sum_i a_i^2\langle m_i^2(0)\rangle}$, for arbitrary vector $\mathbf{a}$. Applying this rule to Eq.~\eqref{eq:evolution} we get
\begin{align}
\rho_{s_1s_2, s_1's_2'}(T) & = \rho_{s_1s_2, s_1's_2'}(0)\exp\left[- \chi^{(\mathrm{resonant})}_{s_1 s_2,s_1' s_2'}(T) \right]\,, \nonumber\\[.3cm]
& =\rho_{s_1s_2, s_1's_2'}(0)\exp\left(- \frac{2T^2\sigma^2  }{\pi^2}\eta_{s_1 s_2,s_1' s_2'}\right)\,,\label{eq:coherence}
\end{align} 
where we defined the {\it resonant attenuation function} $\chi^{(\mathrm{resonant})}_{s_1 s_2,s_1' s_2'}(T)$ associated with coherence $\rho_{s_1s_2, s_1's_2'}$, proportional to {\it resonant attenuation factor} (or simply {\it attenuation factor} for short)
\begin{equation}
\eta_{s_1 s_2,s_1' s_2'} =\bigg|\sum_{q}(s_q-s_q') \rot(-\dphi{q}) \proj\vG{q} \bigg|^2\,.\label{eq:attfc_gen}
\end{equation}

It is vital to recognize that the attenuation factors fall into two categories due to the amount of extractable information on source location. The first category consists of {\it local} attenuation factors,
\begin{align}
\eta_{s_1 s_2,-s_1,s_2}&=  |\rot(-\dphi{1}) \proj\vG{1}|^2 =|\proj\vG{1}|^2\nonumber\\[.1cm]
	& \equiv \,\attfc{11}\,,\\[.25cm]
\eta_{s_1 s_2, s_1, -s_2}&= |\rot(-\dphi{2}) \proj\vG{2}|^2 =|\proj\vG{2}|^2\nonumber\\[.1cm]
	& \equiv \,\attfc{22}\,.
\end{align}
They only include information about one of the qubit-source couplings. Therefore, the same information can be acquired by simply using single-qubit probes (or by relocating a single sensor).

The second, and more interesting category, are the {\it non-local} attenuation factors: the GHZ-type $\eta_{s s, {-s}{-s}}$ and the singlet-type $\eta_{s{-s},{-s} s}$. The corresponding coherences reach the largest possible initial values for maximally entangled GHZ ($\ket{\Phi_{\pm}}$) and singlet ($\ket{\Psi_{\pm}}$) states, but still can be non-zero even for separable states \cite{Szankowski_PRA16}. The key feature of non-local factors is that they not only contain the same information as the local ones, but also provide access to cross-correlation between qubit-source couplings:
\begin{align}
&\eta_{s \,{\pm s}, {{-}s}\, {\mp s}} =\attfc{11}+\attfc{22}\pm 2\,\vG{1}\proj^T\rot(\shift)\proj\vG{2}\nonumber\\[.2cm]
&\phantom{\eta_{\pm}}=\attfc{11}+\attfc{22} \pm 2\bigg[\nonumber\\
&\phantom{\eta_{\pm} =} \vG{1}\proj\vG{2}\cos\shift -\frac{\vB}{|\vB|} \cdot \big( \vG{1} \times \vG{2}\big) \sin\shift \bigg]\nonumber\\[.2cm]
&\equiv \attfc{11}+\attfc{22}\pm 2\,\rechi\cos\phi \mp 2\,\imchi\sin\phi \,.\label{eq:chi_pm}
\end{align}
Here $\shift = \dphi{1} - \dphi{2}$ is the relative phase shift between pulse sequences. Each part of non-local attenuation factor, i.e. the local $\attfc{11}$, $\attfc{22}$, the symmetric (in qubit indexes) $\rechi$ and anti-symmetric $\imchi$, adds unique information about coupling vectors $\mathbf{g}^{(q)}$ and their relative geometry. In contrast, the local attenuation factors depend only on the length of the coupling vectors.

The freedom in choice of pulse sequences allows access to every part of non-local attenuation factor from measurements of a {\it single} coherence. On the other hand, the local attenuation factors can be singled-out by detuning the sequence frequency on the other qubit, thus eliminating a whole contribution from its qubit-source coupling (which also includes the symmetric and anti-symmetric parts of cross-factors, $\rechi$ and $\imchi$). When both $\attfc{11}$ and $\attfc{22}$ have been measured, the symmetric part of cross-factor $\rechi$ is accessed by tuning in both sequences and fixing the relative phase $\phi$ to zero. In these settings the anti-symmetric part is completely suppressed, while the symmetric part is exposed. The anti-symmetric part $\imchi$ is obtained in the \out{analogical}\zi{analogous} manner but for $\phi=\pi/2$.

\subsection{Quasi-static precessing magnetic moment in context of general noise spectroscopy}  \label{sec:spectroscopy}

The formula describing GHZ/singlet-type attenuation factors \eqref{eq:chi_pm} was obtained for a specific relation between filtering properties of SCP sequences and the particular form of time dependence of the source, culminating in Eq.~\eqref{eq:R}. This specialized result can be placed in a broader context of general approach to dynamical decoupling based noise spectroscopy \cite{Szankowski_JPCM17}.

Following the results of \cite{Szankowski_PRA16}, the non-local attenuation functions can be written in terms of overlap integrals of filter functions with so-called self-spectra, $S_{11}(\omega)$ and $S_{22}(\omega)$, and the cross-spectrum, $S_{12}(\omega)$ of the noises experienced by the qubits:
\begin{align}
&\chi_{s{\pm s},{-s}{\mp s}}(T) = \frac{1}{2}\sum_{q}\int |\tilde{f}_T^{(q)}(\omega)|^2 S_{qq}(\omega)\frac{\mathrm{d}\omega}{2\pi} \nonumber\\
&\phantom{\chi_{s{-s}}(T)}\pm \int  \,\mathrm{Re}\Big\{\tilde{f}^{(1)}_T(\omega)\big(\tilde{f}_T^{(2)}(\omega)\big)^*S_{12}(\omega)\Big\}\frac{\mathrm{d}\omega}{2\pi}\,, \label{eq:spectroscopy}
\end{align}
where the attenuation functions $\chi_{s{\pm s},{-s}{\mp s}}$ are evaluated {\it without} assuming that the characteristic frequency of the sequences applied to the qubits, $\omega_q$, is the same as the characteristic frequency of the source, $\omega_s$.
In the case of noise source model discussed here, the spectra are found to be given by combinations of Dirac delta functions,
\begin{align}
S_{qq}(\omega) =& \int \mathrm{d}t\, e^{-i\omega t}\big\langle \big(\vG{q}\cdot\mom(t)\big)\big(\vG{q}\cdot\mom(0)\big)\big\rangle\nonumber\\ 
	=&{\, }2\pi\sigma^2 \,\eta_{qq}\,\frac{\delta(\omega+\omega_s) + \delta(\omega-\omega_s)}{2}\,,\\
S_{1 2}(\omega) =& \int \mathrm{d}t\,e^{-i\omega t} \big\langle \big(\vG{1}\cdot\mom(t)\big)\big(\vG{2}\cdot\mom(0)\big)\big\rangle \label{eq:Sqq}  \\
=&{\,}2\pi\sigma^2\,\rechi \frac{\delta(\omega+\omega_s)+\delta(\omega-\omega_s)}{2}\nonumber\\
	&{}-i2\pi\sigma^2\,\imchi\frac{\delta(\omega+\omega_s)-\delta(\omega-\omega_s)}{2}\,. 
\label{eq:cross-spec}
\end{align}
Note that in this formulation of the problem it is straightforward to take into account the presence of a finite quasi-static broadening $\delta \omega_s$ of the precession frequency of the spins making up the magnetic moment -- one simply has to replace then the delta functions in the expressions above by their finite-width regularizations (e.g.~Gaussians of width $\delta \omega_q$). 

\begin{figure}[tb]
	\includegraphics[width=0.5\textwidth]{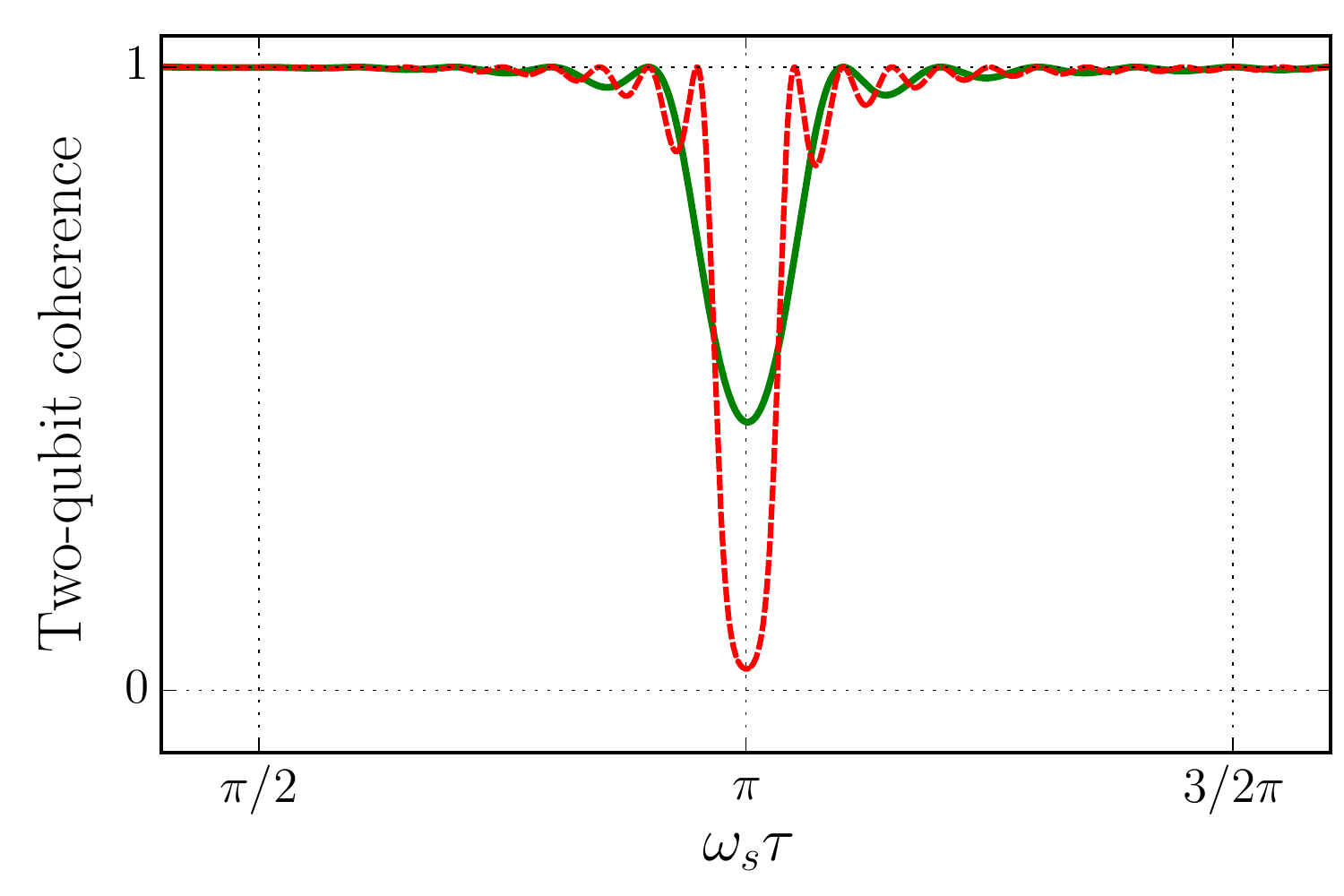}
	\caption{Two-qubit coherence $\rho_{{+\frac{1}{2}}{-\frac{1}{2}},{-\frac{1}{2}}{+\frac{1}{2}}}(n\tau)$ normalized by its initial value plotted for CP sequence with $n\!= \! 20$ as a function of inter-pulse interval $\tau$ (solid, green line), with $\tau_{s} \! =\! \pi/\omega_{s}$ corresponding to the sequence being in resonance with the source. The calculation is performed using Eqs.~(\ref{eq:spectroscopy})-(\ref{eq:cross-spec}). The same for $n\! =\! 40$ pulses, i.e.~for twice as long total sequence duration $T$, is depicted in dashed red line. The minimal value of the coherence follows $e^{-T^2}$ scaling given in Eq.~(\ref{eq:coherence}).
	}\label{fig:dip}
\end{figure}

There are two main points to make here. Firstly, in the case of very narrow bandwidth spectrum considered here, we cannot use the standard approach to dynamical-decoupling based noise spectroscopy, in which the products of $\tilde{f}_T^{(q)}$ in Eq.~(\ref{eq:spectroscopy}) are replaced with Dirac combs that pick out contributions from $S_{ij}(\omega)$ at $\omega \! =\! \omega_q$ and its harmonics. In the present case, in which both the frequency filter, and the noise spectrum, are sharply peaked, the interpretation of the overlap of the spectrum and the filter depends on the relation between the width of the filter maxima ($\sim 1/T$, following from the width of $\mathrm{sinc}$ function in Eq.~(\ref{eq:detune})), and the width of the spectrum, $\delta \omega$. When $\delta \omega \! \ll \! 1/T$, we can indeed treat the spectrum as delta-shaped, and the resulting attenuation factor is proportional to the value of the filter evaluated at $\omega_s$ frequency. We illustrate this in Fig.~\ref{fig:dip}, where we show the two-qubit coherence signal as a function of detuning of $\omega_q$ from $\omega_s$. 
Note that, as discussed in \cite{Szankowski_JPCM17} (see also \cite{Ma_PRAPL16,Ajoy_PNAS17}), the coherence signal calculated using the classical Gaussian noise model, employed throughout the paper, is indistinguishable from the result of a quantum mechanical calculation, in which the magnetic moment is modeled as consisting of noninteracting $M$ spins-$1/2$, each having the same couplings to the two qubits, as long as the number of pulses $n$ fulfills $n \! < \! \omega_{s} / \sqrt{M}A$, where $A$ is the typical magnitude of the spin-qubit coupling. For small $A/\omega_{s}$, i.e.~weak coupling of the  spins of the source to the qubits, and for large $M$, this condition can be fulfilled easily for values of $n$ typically used in noise sensing experiments.

Secondly, in \cite{Szankowski_PRA16} it was argued that when both qubits are coupled to single noise source the cross-spectrum should be a real function. In the model considered here, there is only one source, $\mom$, but Eq.~\eqref{eq:cross-spec} clearly shows that the cross-spectrum is complex. This seemingly paradoxical result is resolved by noting that $\mom(t)$ is a vector quantity and each qubit, in fact, couples to a pair of noises (in the frame where $z\parallel\vB$, those would be $m_x(t)$ and $m_y(t)$). The key point is that the components of noise $\mom(t)$ are correlated in a specific manner, namely
\begin{equation}
m_y(t) = m_x\left (t - \tfrac{\pi}{2}\tfrac{1}{\omega_s} \right ) \,.
\end{equation}
Therefore, the $y$ component of the noise is equal to the {\it retarded} version of the $x$ component. In other words, there is a {\it causal} relation between $m_x$ and $m_y$, and according to \cite{Szankowski_PRA16} such correlations result in non-zero imaginary part of cross-spectrum.

\subsection{Relation to observables}
For the noise source considered here, and for the qubits subjected to dynamical decoupling, the relevant singlet-type and GHZ-type coherences are purely real. Their reconstruction requires then only two tomographic measurement settings:
\begin{equation}
\rho_{s\pm s,-s\mp s}(T) = \mean{\hat S^{(1)}_{x}\hat S^{(2)}_{x}} \mp  \mean{\hat S^{(1)}_{y}\hat S^{(2)}_{y}} \,\, , 
\end{equation}
From practical point of view it is advantageous if the initial value of the modulus of the relevant coherence is as large as possible. As mentioned previously, the maximal initial values for the above coherences are $1/2$ for maximally entangled Bell states, but for separable states (e.g.~both qubits initialized in $+x$ direction) the initial value can be only two times smaller. Two-qubit entanglement is thus not necessary, but its presence increases the amplitude of the signal.

In the next Section we will describe how to infer the position of the source from the values of local and non-local attenuation factors, obtained from measurements of GHZ- or singlet-type coherences. Obviously, any inaccuracy in these measurements will ultimately translate into inaccuracy of the determined position. While the detailed analysis of dominant sources of errors and strategies for mitigation of their influence should be done for each specific physical realization of the discussed setup, let us briefly discuss a few issues related to experimental inaccuracies.
	
The only feature of the initial state that influences the accuracy of the attenuation factor estimation is the initial value of its coherence. As long as this value is large enough to be measured with satisfactory accuracy (e.g. by performing a measurement just after the initialization, with no applied pulses), the imperfections in the qubits' state preparation poses no obstacle for the localization procedure.

More troublesome errors are due to possible imperfections in actual realization of $\pi$-pulses in the DD sequence. For spin qubits controlled by electron spin resonance techniques, the systematic errors of $\pi$-pulses are known to be most dangerous, and for sequences with CP pattern of inter-pulse intervals, their influence is routinely suppressed by appropriately choosing the axes of rotation for subsequent pulses, by e.g.~alternating  between rotations about $x$ and $y$ axes \cite{Gullion_JMR90,deLange_Science10,Staudacher_Science13}. 

The type of error that most strongly affects the localization protocol is the inevitable random spread of measured coherence values, resulting from quantum shot noise (due to averaging over a finite number of projective measurements), and from intrinsic noise of the physical setup used for readout \cite{Degen_RMP17}. The latter typically dominates for currently available spin qubits, and it is especially dangerous, as it leads to variance of measured value of coherence that is independent of this value. In such a case, if the data points of the coherence measurements fell onto curve similar to $n=40$ plot in Fig.~\ref{fig:dip}, then, because of accompanying error bars, the finite dip at $\omega_q=\omega_s$ ($\omega_s\tau=\pi$ in the Figure) could be indistinguishable from zero. Unless the dip can be made shallower by lowering the number of pulses (and thus shortening the duration $T$), such measurement is useless. Since the coherence is an exponential function (see, Eq.\eqref{eq:coherence}), it is arbitrarily close to zero for an infinitely wide range of values of the attenuation factors, thus it becomes impossible to reliably estimate how exactly large is each of them.

If the attenuation factors are not too large, shortening of the pulse sequence allows for coherence measurements that are closer to $n=20$ curve in Fig.~\ref{fig:dip}. Then, it is much easier to recover the values of $\attfc{qq'}$, and their relative error can be estimated to be of the same order as the relative uncertainty of the coherence. On the other hand, in the case of very small attenuation factors (much smaller then the error bars), the relative error of their estimation can be diminished by extending the duration of the sequence instead, so that a shallow dip is made deeper to be more similar to the $n=20$ curve in the Figure.

\section{The localization protocol}  \label{sec:protocol}

\subsection{Assumptions, prerequisites and limitations}

\begin{figure}[t]
	\includegraphics[width=0.5\textwidth]{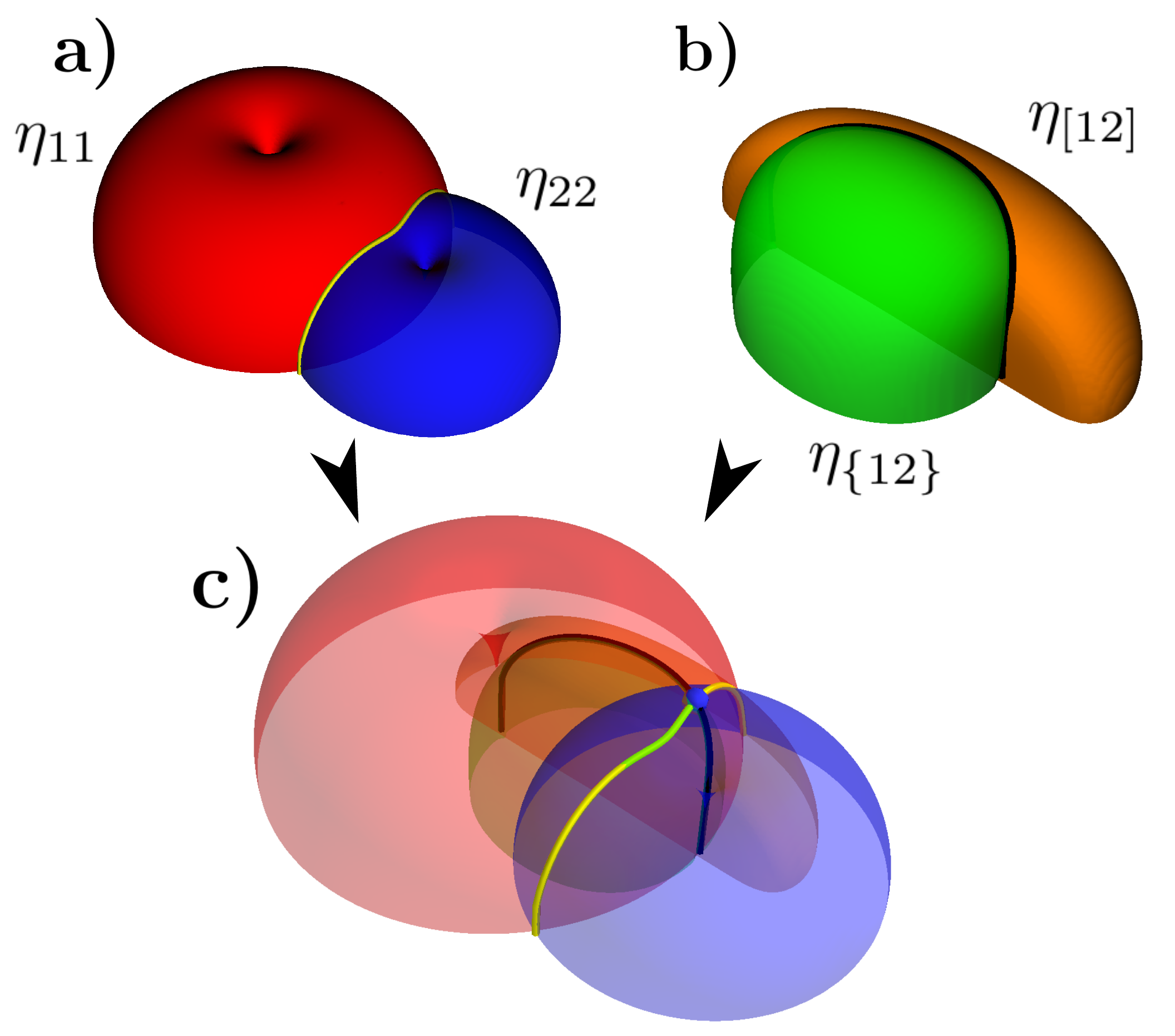}
	\caption{Scheme for the localization of the noise source based on finding the intersections of isosurfaces of (a) $\attfc{11}$ and $\attfc{22}$, (b) $\rechi$ and $\imchi$ treated as functions of qubit-source displacements. In panel (c) one can see how the lines of solutions from (a) and (b) intersect at a single point.  The attenuation factors in this example were calculated for dipole qubit-source coupling of the form given by Eq.~\eqref{eq:dipole_int}. }
	\label{fig:intersect}
\end{figure}

Each part of the attenuation factor $\attfc{11},\attfc{22},\rechi$ and $\imchi$ defines a surface of possible source locations. The ability to apply various pulse sequences grants access to all of those surfaces and consequently to their intersection, which in principle provides enough information to pinpoint the actual source location. Hence, the problem of source localization has been reduced to solving this system of surface equations in order to determine the point (possibly, a set of points) of intersection (see Fig.~\ref{fig:intersect}). Naturally, the applicability of such a method relies on the {\it a priori} knowledge of the qubit-source coupling -- it is impossible to assemble a solvable system of surface equations, unless the form of $\vG{q}(\mathbf{r})$, as a function of qubit-source displacement, is known (which, in consequence, also gives the functional form of $\attfc{qq'}(\mathbf{r}_1,\mathbf{r}_2)$). 

The requirement of having the coupling vectors $\vG{q}(\mathbf{r})$ specified, implicitly assumes that the quantization axis of each qubit is also known. Indeed, coupling vectors originate as an approximation to more general qubit-source coupling, $\hat H_\mathrm{int} = \sum_{q=1,2}\sum_{i=x,y,z} \hat{S}^{(q)}_i \left(\mathbf{n}^{(q)}_i\, \mathbf{G}^{(q)}\, \mom(t)\right)$, where $\mathbf{G}^{(q)}$ is a tensor that depends on qubit-source displacement and $\mathbf{n}^{(q)}_i$ are triplets of orthogonal unit vectors. In the pure dephasing approximation we get
\begin{equation}
\hat{H}_\mathrm{int} \approx \sum_q\hat S^{(q)}_z \left( \mathbf{n}^{(q)}_z\,\mathbf{G}^{(q)}\,\mom(t)\right) \Rightarrow \vG{q} = (\mathbf{G}^{(q)})^T \vnz{q}\,,\label{eq:lin_dep_on_nz}
\end{equation}
hence the coupling vectors $\vG{q}$ are linear functions of respective $\vnz{q}$ -- the unit vector pointing along the $q$th qubit's quantization axis. Therefore, if knowing the qubit-source coupling law is equivalent to knowing the form of tensor $\mathbf{G}^{(q)}(\mathbf{r})$, then in order to derive from it the coupling vectors (and, as a result, the functional form of attenuation factors) one also needs the quantization axes $\vnz{q}$.

The multiplicative factors in Eq.~\eqref{eq:coherence}, that include the duration of pulse sequence $T$ and the rms of the moment's magnitude, $\sigma$, must be known {\it a priori} with satisfactory accuracy. In the approach presented here it is impossible to obtain them as part of solution to the system of surface equations. The reason is as follows. If the qubit-source interaction is a homogeneous function, the rescaling of length units leads to 
\begin{equation}
\attfc{qq'}(s\mathbf{r}_1 , s\mathbf{r}_2) = s^\beta \attfc{qq'}(\mathbf{r}_1,\mathbf{r}_2)\,,
\end{equation}
For example, in case of dipole qubit-source coupling (see, Eq~\eqref{eq:dipole_int}), $\beta = -6$. The aforementioned multiplicative factors effectively play the same role as the scale $s$ and because of that, they cannot be treated as additional unknowns.

Finally, note that because of the translational symmetry of the qubit-source interaction it is only possible to infer the {\it relative} position of the source in respect to the sensor qubits (hence, the set of surface equations is to be solved for qubit-source displacements). Therefore, the accuracy of the localization method is limited by the {\it a priori} knowledge of the sensor position, which involves either knowing the position of both sensor qubits, or just one of them. The protocols for both of those cases are discussed below.

\subsection{Localization with both sensor qubits located}\label{sec:known_d}

First we discuss the case when the positions of both sensor qubits are known. Since each qubit-source displacement vector point towards the same source position, one of them is uniquely determined by the other, say $\vR{2}= \vR{1}+\mathbf{d}$, where $\mathbf{d}$ is a known qubit-qubit displacement vector. Therefore, the localization of the source is reduced to determining three vector components, e.g. $r^{(1)}_x$, $r^{(1)}_y$, $r^{(1)}_z$. As described above, the measurements of coherence decay give access to at most four surface equations:
\begin{equation}
\left\{\begin{array}{rcl}
\attfc{11}(\mathbf{r},\mathbf{r}+\mathbf{d})\big|_{\mathbf{r}=\vR{1}} &=& \attfc{11}^{(\text{ex})}\\[.2cm]
\attfc{22}(\mathbf{r},\mathbf{r}+\mathbf{d})\big|_{\mathbf{r}=\vR{1}}  &=& \attfc{22}^{(\text{ex})}\\[.2cm]
\rechi(\mathbf{r},\mathbf{r}+\mathbf{d})\big|_{\mathbf{r}=\vR{1}}  &=& \rechi^{(\text{ex})}\\[.2cm]
\imchi(\mathbf{r},\mathbf{r}+\mathbf{d})\big|_{\mathbf{r}=\vR{1}}  &=& \imchi^{(\text{ex})}\\
\end{array}\right.\,,\label{eq:all_is_known_set}
\end{equation}
where the superscript $(\text{ex})$ refers to the measured values of corresponding attenuation factor. This system is overdetermined since there are only three unknowns, hence one can construct four sets of three equations. This overabundance of equations can be very beneficial. Due to symmetries of the qubit-source interaction, each set can yield a whole ensemble of possible source locations. However, all four equations must be consistent (since the values of attenuation factors were measured for a particular position of the source), and only solutions shared by all the sets are valid. This ability to cross-check improves the confidence level of the method.

However, the independence of equations in each set is not guaranteed and it should be verified. This can be accomplished by evaluating the Jacobi determinant of a vector function $\mathbf{h}(\mathbf{r}) =(h_1(\mathbf{r}),h_2(\mathbf{r}),h_3(\mathbf{r}))$
\begin{equation}
J(\mathbf{r})=\left|\frac{\partial(h_1,h_2,h_3)}{\partial(r_x,r_y,r_z)} \right|
=\text{det}\left(\begin{array}{ccc}
\frac{\partial h_1}{\partial r_x} & \frac{\partial h_1}{\partial r_y} & \frac{\partial h_1}{\partial r_z} \\
\frac{\partial h_2}{\partial r_x} & \frac{\partial h_2}{\partial r_y} & \frac{\partial h_2}{\partial r_z} \\
\frac{\partial h_3}{\partial r_x} & \frac{\partial h_3}{\partial r_y} & \frac{\partial h_3}{\partial r_z} \\
\end{array}\right)
\,,
\end{equation}
where $h_i$ stand for the left-hand-side of a given set of surface equations, e.g. for a set composed of first three equations of (\ref{eq:all_is_known_set}) one would get $h_1(\mathbf{r}) = \attfc{11}(\mathbf{r},\mathbf{r}+\mathbf{d})$, $h_2(\mathbf{r}) =\attfc{22}(\mathbf{r},\mathbf{r}+\mathbf{d})$ and $h_3(\mathbf{r}) =\rechi(\mathbf{r},\mathbf{r}+\mathbf{d})$. If the Jacobian at point $\mathbf{r}$ is non-zero, then the vector function $\mathbf{h}(\mathbf{r})$ is invertible in the neighborhood of that point. This means that locally there is a one-to-one correspondence between selected attenuation factors and the position of the source, or in other words, the solutions to the system of equations is a set of isolated points. If the Jacobian vanishes in a certain volume (or on a surface, or a curve) then the equations are satisfied by a whole continuum of points. Such a situation occurs when the qubit-source coupling is highly symmetric, so that the intersection of surfaces defined by the attenuation factors is a 1D curve or even a 2D surface itself, instead of a set of points. This happens, for example, for isotropic coupling $\vG{q}(\mathbf{r}) = \vG{q}(|\mathbf{r}|)$. As a side note, if the multiplicative factors mentioned previously were treated as unknown in the system of four surface equations, the Jacobi determinant would vanish identically due to unit-rescaling feature of attenuation factors described above.

In practice, there is no guarantee that the cross-check between sets of solutions obtained from all four available equation systems will be sufficient to eliminate all of them, except for the one corresponding to the true location of the source. A natural work-around for this issue is to supplement the original equations with additional ones, that would yield even more solutions to cross-check with. For example, a set of four new surface equations can be generated by re-orienting the control field $\vB$ that alters the plane on which couplings $\vG{q}$ are being projected (see, Eq.~\eqref{eq:R}). This effectively changes the functional form of $\proj\vG{q}$, and consequently produces a new set of four equations. By design, these new equations share the same unknown with the original set, because the positions of the source and the sensor were not changed. Hence, with only one alteration of the magnetic field direction, the method produces eight consistent equations in total, and there are ${{8}\choose{3}} = 56$ ways to choose from among them a system of three. (Of course, each set should be verified by inspecting an appropriate Jacobian determinant.) This relatively large number of sets of equation systems offers plenty opportunities to cross-check solutions, and narrow down possible source locations even further. In case when this still proves not to be enough, an additional sets of equations can be added by considering more settings of the magnetic field.

The results of numerical simulations testing the protocol are presented in Appendix~\ref{apx:numerical_tests}.

\subsection{Localization with only one sensor qubit located}\label{sec:unknown_d}

If the qubit-qubit displacement $\mathbf{d}$ is unknown then both $\vR{1}$ and $\vR{2}$ are to be treated as independent unknowns and both are needed to localize the source. In principle it requires at least six independent surface equations to form a solvable system. As discussed previously, a single source-sensor setup can produce at most four equations. Therefore, the method for generating additional sets of surface equations by altering the magnetic field $\vB$ introduced in previous section, must be employed in this case. The minimal required number of equations can be obtained in such a way with only one additional setting of the field.

\section{Localization in case of unknown coupling law -- the symmetry sensor}  \label{sec:symmetry}

When the qubit-source coupling law is unknown, the notion of localizing the source by finding the intersection point of attenuation factor-defined surfaces has to be abandoned.  Instead, we propose a protocol where the sensor is moved and read out at various locations, while the source remains static. In this way the map of attenuation factors for a selection of source-sensor arrangements is created. The idea is that this map has some universal, coupling independent feature that singles out a specific arrangement, allowing to extrapolate the location of the source from the known position of the sensor. Our choice is to look for zero of the attenuation factor map. This is a particularly pragmatic option from the experimental point of view, because vanishing of $\attfc{s s, {-s}{-s}}$ or $\attfc{s {-s},{-s} s}$ would be observed as a significant increase of the coherence time.

Assuming that the coupling law between each of the sensor qubits and the source are the same, the coupling vectors can be written as a vector functions
\begin{equation}
\vG{q}  = \mathbf{g}(\vR{q},\vnz{q})\,,
\end{equation}
where $\vR{q}$ is the displacement between source and qubit $q$, and $\vnz{q}$ is the unit vector along qubit's quantization axis (see, Eq.~(\ref{eq:Hint})). Substituting this form of couplings into Eq.~(\ref{eq:attfc_gen}), the decay of GHZ- and singlet-type coherences reads 
\begin{equation}
\attfc{s {\pm s}, {-s} {\mp s}} = \left| \proj\mathbf{g}(\vR{1},\vnz{1}) \pm \proj\mathbf{g}(\vR{2},\vnz{2})\right|^2\,.\label{eq:chi_no_pulses}
\end{equation}
(Here we consider the case when the identical pulse sequences are applied to each qubit, so that $\shift=0$). One situation in which the above attenuation factor vanishes is when the source and the sensor happen to be arranged in such a way that the projected coupling vector cancel each other out. At first glance, it might seem impossible for such an arrangement to exist. Even if quantization axes are the same ($\vnz{1}=\vnz{2}$), the qubit-source displacements are always different because $\vR{2} =\vR{1}+\mathbf{d}$ and $\mathbf{d}\neq 0$. However, it is reasonable to expect a certain degree of symmetry in qubit-source coupling, so that there is a special source-sensor arrangement for which $\attfc{s s, {-s}{-s}}$ or $\attfc{s{-s},{-s}s}$ would vanish anyway. Below we present a prescription for sensor setups, that single out a well defined family of source-sensor arrangements for which the coupling vectors cancel exactly, causing the vanishing of the attenuation factors in Eq.~(\ref{eq:chi_no_pulses}).

Let the sensor setup satisfy the following prerequisites: i) the applied field $\vB$ is set to be perpendicular to the qubit-qubit displacement vector $\mathbf{d}$, i.e. $\vB\perp\mathbf{d}$, ii) the qubits' quantization axes are related through $\pi$-rotation around $\vB$,
\begin{equation}
\vnz{2} = \pm\rot(\pi) \vnz{1}\,.\label{eq:asumpt}
\end{equation}
(see Eq.~\eqref{eq:R} for the definition of rotation matrix $\rot$.) Depending on the sign choice in \eqref{eq:asumpt}, we shall refer to such a sensor setup as the {\it upper} or {\it lower sign Null setup}. For example, the simplest realization of the upper sign Null setup is obtained when both qubit axes are parallel to $\vB$, while the lower sign Null setup requires that one of the axes is anti-parallel (and of course, $\vB\perp\mathbf{d}$ must also be satisfied.) 

\begin{figure}[htb]
	\includegraphics[width=0.5\textwidth]{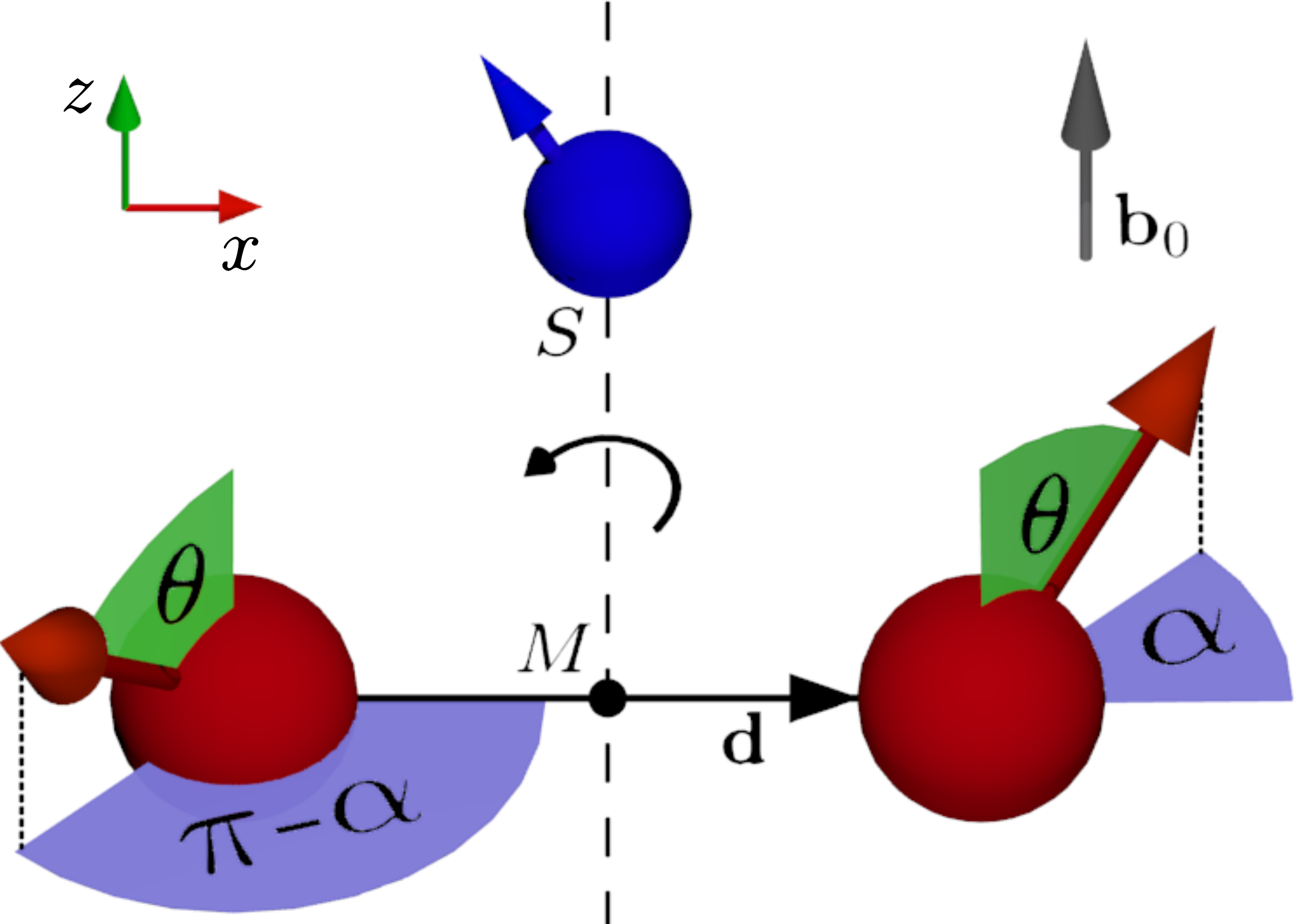}
	\caption{The schematic diagram of the GHZ {\it Null arrangement}: the source-sensor arrangement corresponding to vanishing GHZ-type attenuation factor $\attfc{s s,{-s}{-s}}$ for arbitrary qubit-source coupling law. The source may be located anywhere on the {\it Null axis} $MS$, which is i) perpendicular to qubit-qubit displacement vector $\mathbf{d}$, ii) passes through $M$, the midpoint of $\mathbf{d}$. The sensor has been initialized according to the {\it upper sign Null setup}, which means that the field $\vB$ is perpendicular to $\mathbf{d}$, and the quantization axis of qubit $1$, $\vnz{1}=(\sin\theta\cos(\alpha-\pi),\sin\theta\sin(\alpha-\pi),\cos\theta)=(-\sin\theta\cos\alpha,-\sin\theta\sin\alpha,\cos\theta)$, is such, that when it is rotated by angle $\pi$ around $\vB$ it exactly matches the quantization axis of qubit $2$, $\vnz{2}=(\sin\theta\cos\alpha,\sin\theta\sin\alpha,\cos\theta)=\rot(\pi)\vnz{1}$. (see, Eq.~\eqref{eq:asumpt}). }\label{fig:config}
\end{figure}

If the sensor is in upper (lower) sign Null setup, the projected coupling vectors in Eq.~\eqref{eq:chi_no_pulses} become opposite (identical) to each other, when the sensor is placed in such a way that the source is situated anywhere on the {\it Null axis} --- the line parallel to $\vB$ and passing through the midpoint of $\mathbf{d}$. We shall refer to such a source-sensor arrangement, which causes the GHZ-type (singlet-type) attenuation factor to vanish, as the {\it GHZ (singlet) Null arrangement}. The schematic diagram depicting the GHZ Null arrangement is presented in Fig.~\ref{fig:config}.

Here we demonstrate the proof of this proposition. In order to simplify the presentation, let the field $\vB$ define the $z$-axis, while the qubit-qubit displacement $\mathbf{d}$ the $x$-axis of the reference frame with the origin at the midpoint of $\mathbf{d}$. Suppose the qubits' axes satisfy (\ref{eq:asumpt}) with the {\it upper} sign and the source is located at $(0,0,z)$, corresponding to GHZ Null arrangement. Then, the qubit-source displacements are given by $\vR{1} = (|\mathbf d|/2 , 0 , z)$ and $\vR{2} = (-|\mathbf d|/2,0,z)$. Note that in such a case, not only $\vnz{1}$ and $\vnz{2}$ are related through the $\pi$-rotation, but also $\vR{2} = \rot(\pi)\vR{1}$. Therefore, the axis and the displacement of the second qubit are identical to those of the first one, but viewed in the rotated reference frame. It follows from the general transformation rules for vector functions, that $\mathbf{g}$ evaluated at $\vnz{1}$ and $\vR{1}$ equals $\mathbf{g}$ viewed in the rotated frame and evaluated at the transformed inputs, i.e.
\begin{align}
\nonumber
&\mathbf{g}(\vR{2},\vnz{2}) =
	\mathbf{g}(\rot(\pi)\vR{1},\rot(\pi)\vnz{1} )\\
&=\rot(\pi)\,\mathbf{g}(\vR{1},\vnz{1})\,.
\end{align}
Since the rotation axis is set by the magnetic field, the components unaffected by $\rot$ are removed from $\proj\vG{q}$ by the projection, and the GHZ-type attenuation factors reads
\begin{align}
&\attfc{ss,{-s}{-s}} =| \proj\vG{1}+\proj\vG{2}|^2
	=| \proj(\mathbf{I} + \rot(\pi))\vG{1} |^2\nonumber\\
&=\left| \left[\left(\begin{array}{ccc}
				1 & 0 & 0\\0 & 1 & 0\\ 0 & 0 & 0\\ \end{array}\right)
		+\left(\begin{array}{ccc}
					-1 & 0 & 0\\ 0 & -1 & 0 \\ 0&0&0 \\ \end{array}\right)\right]
			\vG{1}\right|^2=0\,.
\end{align}
As we can see, it vanishes identically. 

In the same settings the singlet-type attenuation factor is instead enhanced (see, Eq.~(\ref{eq:chi_no_pulses})). However, the application of pulse sequences that are appropriately time shifted, so that $\phi\! =\! \pi$ effectively transforms $\attfc{s{-s},{-s}s}$ into $\attfc{s s,{-s}{-s}}$, which makes it disappear anyway. 

Alternatively, $\attfc{s{-s},{-s}s}$ vanishes in the singlet Null arrangement, when an additional property of $\mathbf{g}$ is exploited, namely, its linear dependence on the qubit axis (see, Eq~\eqref{eq:lin_dep_on_nz}).  
Now, assuming that qubits' axes satisfy the {\it lower} sign Null setup we can repeat the previous steps, with an adjustment
\begin{align}
&\mathbf{g}(\vR{2},\vnz{2}) = \rot(\pi)\mathbf{g}(\vR{1},-\vnz{1})\nonumber\\
&=-\rot(\pi)\mathbf{g}(\vR{1},\vnz{1})\,,
\end{align}
so that
\begin{align}
\attfc{s{-s},{-s}s} &=| \proj[\mathbf{I} -(-\rot(\pi))]\,\vG{1} |^2 \nonumber\\
	&= |\proj(\mathbf{I} +\rot(\pi))\,\vG{1} |^2 =0\,.
\end{align}
Which concludes the proof.

The fact that the Null arrangement is achieved whenever the source is situated {\it anywhere} on the Null axis means that simply finding a zero of attenuation factor map does not allow to pinpoint the location of the source. Instead, it narrows down all possible locations of the source to a line. This ambiguity can be resolved by performing another attenuation factor mapping using sensor with qubit-qubit displacement vector tilted with respect to the original. The tilted vector $\mathbf{d}$ requires modified Null setup, which in turn alter the corresponding Null arrangements, and redefines the Null axis. The new line of possible locations obtained in such a way can intersect with the original line only in one point, thus revealing the location of the source. This extension can be achieved either by bringing in the second pair of qubits, or by rotating the original sensor, or alternatively, by rotating the probed sample.

It is important to recognize that, even for sensor in Null setup, the Null arrangements described above are not necessary the only source-sensor arrangements producing zeros in the attenuation factor map. Indeed, if the source-qubit coupling law is highly symmetric, the Null axis might be downgraded to a Null plane. For example, in case of isotropic coupling $\vG{q} = \mathbf{g}(|\vR{q}|,\vnz{q})$, the attenuation factor vanishes due to cancellation of the coupling vectors whenever the source is situated anywhere in the plane that contains the Null axis and is normal to qubit-qubit displacement vector $\mathbf{d}$. Then, in order to reveal the location of the source, three attenuation factor mappings are required, in comparison to only two in the case of less symmetric coupling law. The zeros of attenuation factor map can also be ``accidental'': it is possible that due to specific properties of the coupling law there exist non-Null source-sensor arrangements for which the coupling vectors $\proj\vG{q}$ cancel out, or even simultaneously vanish. Such accidental zeros of the map cannot be used for extrapolation of the source localization. However, those zeros can be easily distinguished from the zeros due to Null arrangements. If a given zero results from the sensor being placed in the Null arrangement with the source, then the Null arrangement would also be achieved when the sensor is moved anywhere along $\vB$ (which corresponds to situating the source along the Null axis). Therefore, if the zero of the map is not accompanied by a line of other zeros characteristic for the family of Null arrangements, it must be accidental.

\begin{figure}[htb]
	\includegraphics[width=0.5\textwidth]{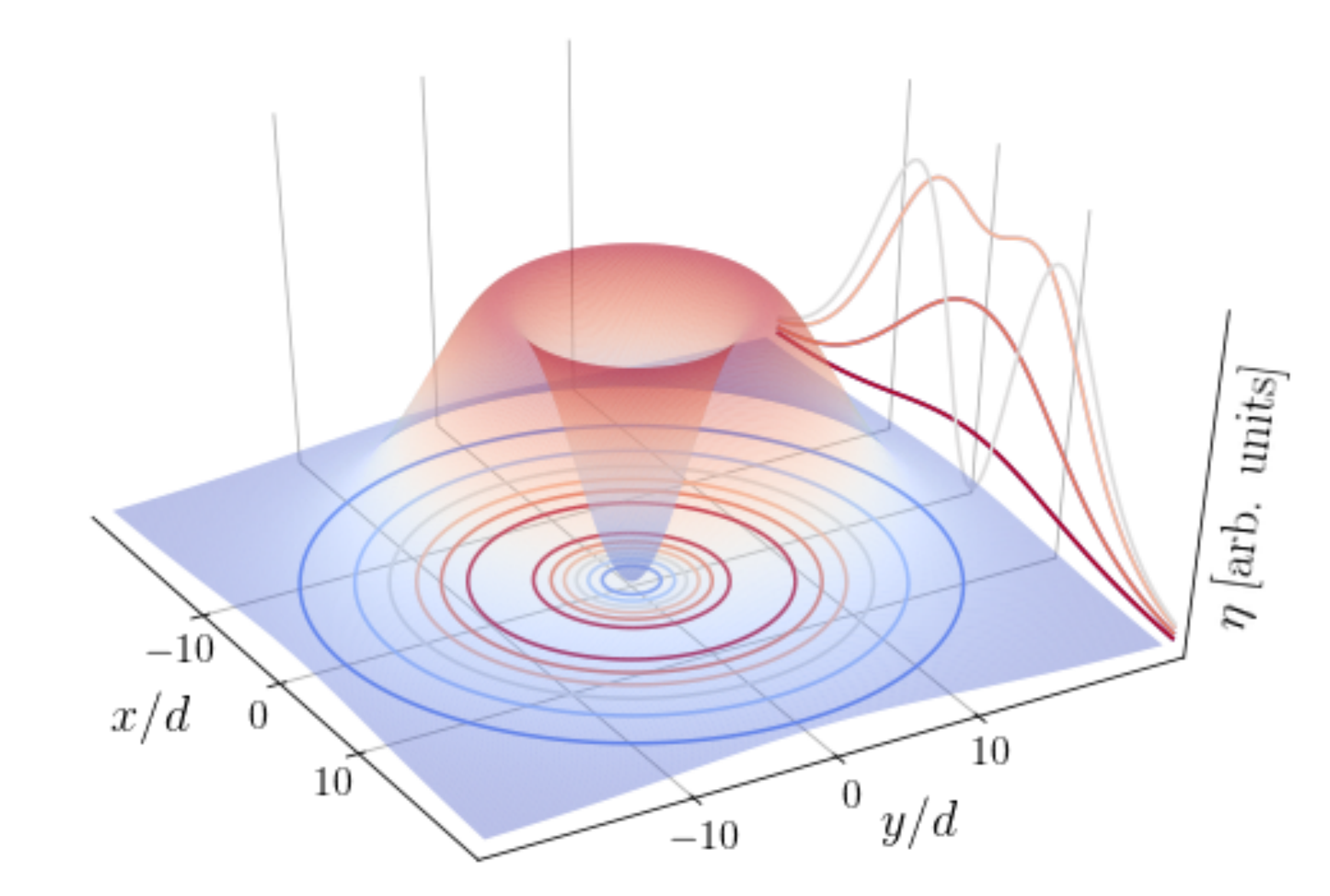}
	\caption{
	An example of the restriction of GHZ-type attenuation factor map to a plane of sensor positions for the case of dipole qubit-source coupling (see, Eq~\eqref{eq:dipole_int}). The qubit-qubit displacement vector $\mathbf{d}$ is set parallel to $x$-axis. The magnetic filed $\vB$ and quantization axes $\vnz{q}$ are aligned with the $z$-axis, so that the sensor is in the upper sign Null setup. The source is placed at fixed location $\mathbf{r}^{(s)} = (0,0,10|\mathbf{d}|)$. The attenuation factor map was created for arrangements where the sensor was confined to $xy$-plane at $z=0$. When the sensor is placed directly below the source ($x=y=0$), or equivalently the source is situated on the Null axis, the attenuation factor vanishes. The shape of the crater rim, where the attenuation factor reaches local maximum of $\sim (r^{(s)}_z)^{-6}$, is approximately circular with diameter $\sim r_z^{(s)}$. The crater becomes more elliptical when $r_z^{(s)}$ is comparable with inter-qubit distance $|\mathbf{d}|$.
	}\label{fig:volcano}
\end{figure}

If we exclude the possibility of accidental zeros and the existence of Null planes, the measured attenuation factors would be non-zero, except for the family of Null source-sensor arrangements. Nevertheless, one should expect for the landscape of attenuation factor maps to possess general features independent of the  form of the qubit-source coupling. When the sensor is far away from the source, the coupling to each qubit is small due to natural tendency for interaction strength to diminish with distance. As the sensor is brought closer to the source, the distance shrinks, and the attenuation factors increase until they reach (local) maximum. Moving the sensor even closer, the attenuation factor drops until it disappears as the sensor is brought closer to the Null arrangement, and the source is situated in the vicinity of the Null axis. Therefore, one should expect a volcano-shaped attenuation factor map, with the Null arrangement at the center of the crater. We illustrate this point with an example of dipole qubit-source interaction presented in Fig.~\ref{fig:volcano}. 

Since the features of this landscape are determined by the behavior of the attenuation factors alone, they are independent of the multiplicative factors present in Eq~\eqref{eq:coherence}. This means that, in contrast to the localization technique described previously, this approach can be implemented in case when $\sigma$ is unknown.

\section{Conclusion}

We have presented proposals of two methods for localization of a magnetic moment precessing with well defined frequency by a sensor composed of two dynamically decoupled qubits. The first method utilized {\it a priori} knowledge of the functional form of qubit-source interaction (e.g.~dipolar coupling) to infer the position of the source by finding the intersection of surfaces defined by the decay of the two-qubit coherence. The second method, which can be deployed even if the coupling law is unknown, requires the ability to move the sensor. By measuring the decay of the coherence at various sensor location the map of attenuation factor is created. For a properly set up sensor, the zero of this map defines the axis of possible source locations. 

The presented methods rely on the ability to apply two dynamical decoupling sequences of pulses to the two qubits, and thus they are examples of practical applications of recently discussed multi-qubit DD protocols \cite{Szankowski_PRA16,Paz_NJP16,Paz_PRA17}. Note that application of DD sequence is routinely done for a single qubit in nanoscale magnetometry experiments with NV centers, and applying two distinct sequences to two nearby qubits should be feasible as long as their energy splittings are distinct, e.g. due to presence of magnetic field gradient or their quantization axes are non-parallel. 
They additionally require tomographic reconstruction (using only two combinations of measurement settings) of one of two-qubit coherences. The coherences of interest are nonzero for separable two-qubit states, so creation of two-qubit entanglement, while helpful since its presence maximizes the values of the coherences, is not necessary.

\section*{Acknowledgements}
This work is supported by funds of Polish National Science Center (NCN), grants no.~DEC-2012/07/B/ST3/03616 and DEC-2015/19/B/ST3/03152. We would like thank Damian Kwiatkowski for his comments on the manuscript.

\appendix\section{Numerical tests of triangulation protocol}\label{apx:numerical_tests}
In order to test the usefulness of the triangulation protocol we have implemented appropriate simulations for the dipole qubit-source coupling (see, Eq.~\eqref{eq:dipole_int}). The first sensor qubit was placed at origin of the reference frame, which coincided with the center of a cubic box with an edge length of  $20|\mathbf{d}|$, where $\mathbf{d}$ is the qubit-qubit displacement vector. The second qubit was placed at $(|\mathbf{d}|,0,0)$. While the position of the sensor was always fixed, the source could be placed anywhere in the upper-half of the box ($z>0$).  

\begin{figure}[tb]
	\includegraphics[width=0.5\textwidth]{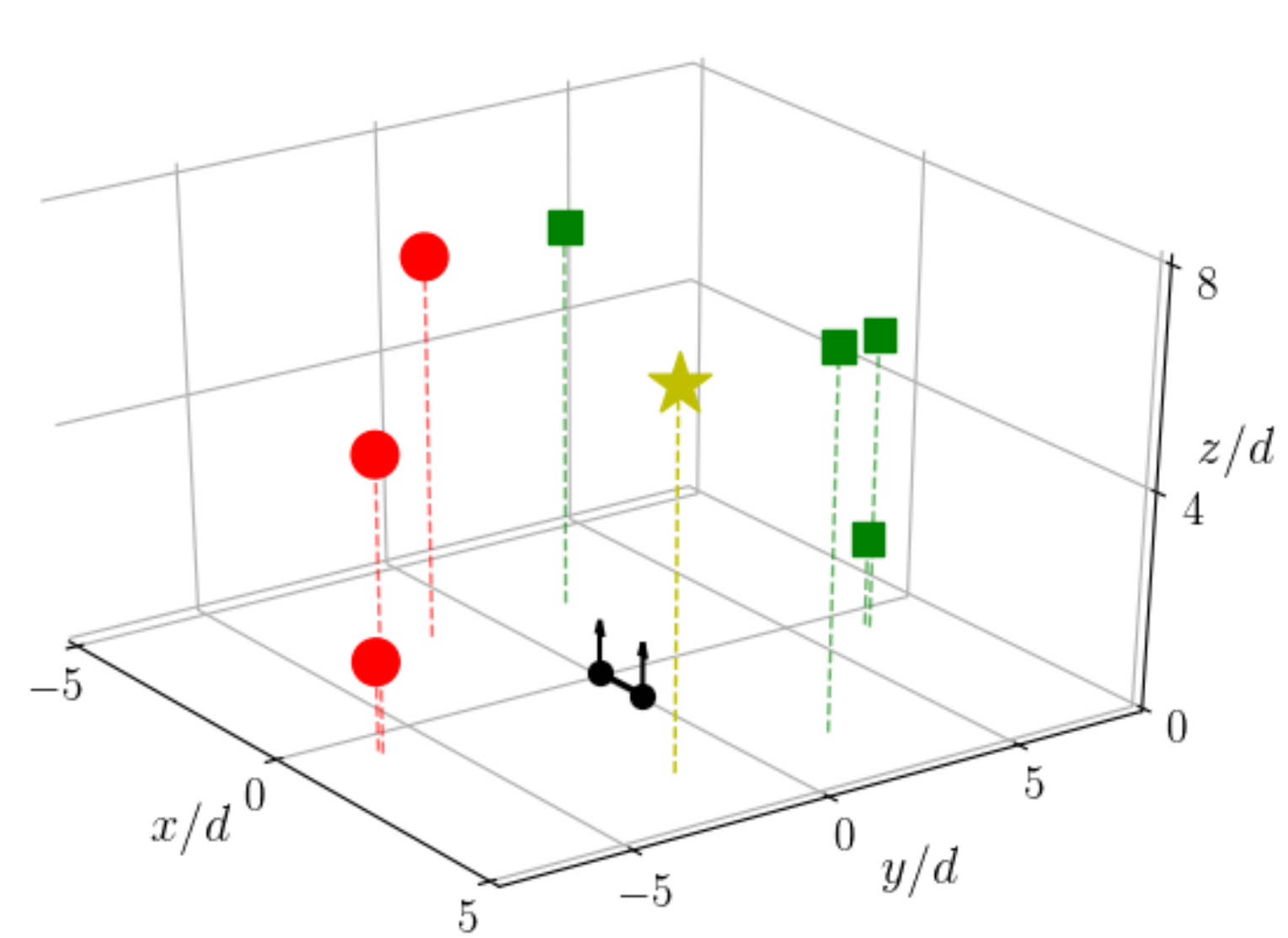}
	\caption{The depiction of the solution cross-checking process in a typical instance of the localization protocol simulation. The figure shows only the upper-half of the box. The schematic diagram of the sensor is drawn for reference (the qubit-qubit displacement $\mathbf{d}$ is in scale). The symbols (squares, circles and the star) indicate the numerical solutions to a single system of three surface equations formed for $\vB/|\vB|=(0,0,1)$ magnetic field orientation.  The green squares are the solutions exclusive to this system of equations. The non-square symbols (red circles and the star) indicate solutions that where positively cross-checked with the sets of solutions found to the remaining three systems of equations, i.e. solutions shared by all four systems possible to form with single magnetic field orientation. Finally, the golden star is the only solution that is shared among the original systems of equations and the additional four systems formed for field orientation $\vB'/|\vB'|=(1/\sqrt 2,0,1/\sqrt 2)$. This singled out position coincides with the real source location.}
	\label{fig:sol_cross-check}
\end{figure}
\begin{figure}[tb]
	\includegraphics[width=0.5\textwidth]{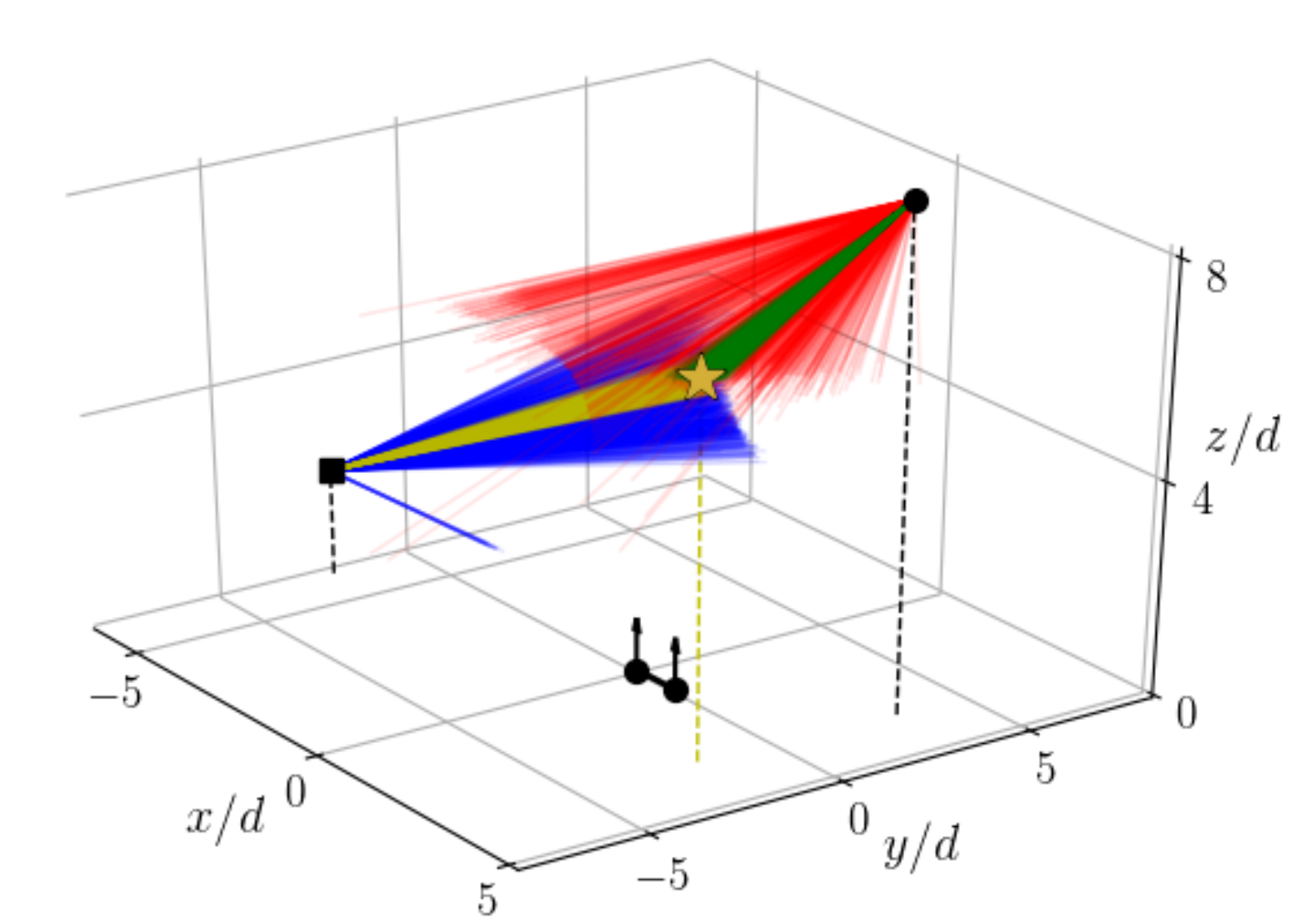}
	\caption{An illustration of the effects of measurement errors on the localization protocol. The golden star marks the real source location, and the schematic diagram of the sensor is drawn for reference (the qubit-qubit displacement $\mathbf{d}$ is in scale). The rays connect the starting points---marked with square for one setting of the magnetic field and circle for another---and the solutions found by the algorithm. The input data for each run in a form of attenuation factors values, was distorted by a random noise. The simulated noise affecting the measured $\attfc{qq'}$ had Gaussian distribution with zero average and dispersion $\sigma_{\attfc{qq'}}=0.1\times\attfc{qq'}$ (blue and red rays) or $\sigma_{\attfc{qq'}}=0.01\times\attfc{qq'}$ (yellow and green rays). In the case of dipole source-qubit coupling, the spread of the input data resulted in a arc-shaped ``cloud'' of outputs. Its orientation clearly depends on the magnet field settings.}
	\label{fig:uncertain}
\end{figure}

 We used Levenberg-Marquardt algorithm implemented in Python library SciPy as our method for solving systems of surface equations. The tests of the localization method were carried out according to the following scheme:

\begin{enumerate}
	\item First we drew 100 instances of the source position within the upper-half of the box from a uniform distribution
	\item In order to simulate the measured values of the attenuation factors for two orientations of magnetic field, $\vB/|\vB|=(0,0,1)$ and $\vB'/|\vB'|=(1/\sqrt 2,0,1/\sqrt 2)$, we substituted the drawn positions into known qubits-source coupling law. Thus, we obtained two sets of attenuation factors for each setting of the field: $\{\attfc{11}^{(\alpha)},\attfc{22}^{(\alpha)}, \rechi^{(\alpha)},\imchi^{(\alpha)}\}$ and $\{\attfc{11}'^{(\alpha)},\attfc{22}'^{(\alpha)}, \rechi'^{(\alpha)},\imchi'^{(\alpha)}\}$. The superscript $\alpha$ indicates the instance of the source position, and it runs from $1$ to $100$.
	\item For each simulated experiment (i.e. each $\alpha$) we formed two sets---one for each magnetic field setting---of four systems of triplets of surface equations, with $\attfc{qq'}^{(\alpha)}$ or $\attfc{qq'}'^{(\alpha)}$ calculated in the previous step, substituted for the right-hand-side of each equation. Therefore, for each instance of the source position, we had eight systems of equations to solve.
	\item Aside from the system of equations, the algorithm requires a starting value for the unknown. For each system of equations, the algorithm was run 100 times with the initial values drawn at random from the uniform distribution inside the box. The solution obtained in each of the runs was kept only if the algorithm converged before it reached the maximal number of iterations set to 10 000. A run was considered to be converged, when the relative difference between results of consecutive iterations was smaller than $10^{-30}$, and the solution was confined within the upper-half of the box. This means that the algorithm was executed $100\times2\times4\times100=80\,000$ times (about $25$\% of the obtained solutions were discarded due to convergence criterion), which took no more then few minutes.
	\item For each $\alpha$, we cross-checked the sets of solutions to all eight systems of equations found by the algorithm. We kept only those solutions which where shared by each set. (We considered two solutions to be the same when the relative distance between them was smaller then $10^{-16}$.) 
\end{enumerate}

For each instance of the source position, the cross-check between solutions to systems of surface equations formed for two orientations of the magnetic field, provided enough information to pin-point the true source location. Figure~\ref{fig:sol_cross-check} depicts the course of a typical experiment simulation: it illustrates how cross-checking with additional systems of surface equations leads to elimination of possible solutions that do not correspond to actual position of the source.

In addition, we have tested how the results returned by the algorithm are modified by the addition of measurement errors. For this purpose we have fixed the source location and two starting values of the algorithm (one for each magnetic field setting). Then, for each field orientation, we have performed $2000$ runs in a mode where the algorithm was looking for a simultaneous solution to all four surface equations (effectively yielding the same results as a cross-check between the solutions of four systems of three equations). In order to simulate the measurement uncertainty, in each run we modified the exact values of the attenuation factors, $\attfc{qq'}\to\attfc{qq'}+\Delta\attfc{qq'}$, where $\Delta\attfc{qq'}$ was drawn at random from zero average Gaussian distribution. One half of the runs were performed with the dispersion of the distribution set to $10$\% of the value of the perturbed attenuation factor, and the other half to only $1$\% of this value. A typical result (as compared to different choices of the source location and/or starting values) is illustrated in Fig.~\ref{fig:uncertain}. As expected, the spread in the input data (the errors of the attenuation factors measurements) causes a spread of the numerical solutions. The shape of the ``cloud'' of solutions, obviously, depends on the properties of the source-qubit coupling law. In our case, the solutions are arranged along a long and narrow arc. As we can see on the figure, the orientation of this arc depends on the orientation of the magnetic field. However, the spread of results obtained individually for each magnetic field setting can be effectively narrowed by determining the overlap between the two arcs. Indeed, the characteristic size of this overlap of cross-checked solutions is given by the arcs' widths, which are much smaller then their lengths.


%

\end{document}